\documentclass[10pt,aps,prc,twocolumn,superscriptaddress,floatfix,nofootinbib]{revtex4-1}
\usepackage{times}
\usepackage[T1]{fontenc}

\usepackage{amsmath,amssymb,mathrsfs}
\usepackage{bm}
\usepackage{ascmac}
\usepackage{geometry}
\usepackage{graphicx}

\def\0\\{\nonumber\\}
\def\bs#1{\boldsymbol{#1}}

\def\fs#1{{\footnotesize #1}}

\makeatletter
\newcommand\footnoteref[1]{\protected@xdef\@thefnmark{\ref{#1}}\@footnotemark}
\makeatother

\begin{document}

\title{
Quantal diffusion approach for multinucleon transfer processes in the $^\mathbf{58,64}$Ni\,+\,$^\mathbf{208}$Pb reactions:
Toward the production of unknown neutron-rich nuclei
}

\author{Kazuyuki Sekizawa}
\email[]{sekizawa@phys.sc.niigata-u.ac.jp}
\affiliation{Center for Transdisciplinary Research, Institute for Research Promotion, Niigata University, Niigata 950-2181, Japan}
\affiliation{Division of Nuclear Physics, Center for Computational Sciences, University of Tsukuba, Ibaraki 305-8577, Japan}

\author{Sakir Ayik}
\email[]{ayik@tntech.edu}
\affiliation{Physics Department, Tennessee Technological University, Cookeville, Tennessee 38505, USA}

\date{July 29, 2020}

\begin{abstract}
\begin{description}
\item[Background]
In recent years, substantial efforts have been made for the study of multinucleon transfer reactions at
energies around the Coulomb barrier both experimentally and theoretically, aiming at the production
of unknown neutron-rich heavy nuclei. It is crucial to provide reliable theoretical predictions based
on microscopic theories with sufficient predictive power.

\item[Purpose]
This paper aims to clarify the applicability of the quantal diffusion approach based on
the stochastic mean-field (SMF) theory for multinucleon transfer processes. Isotope
production cross sections are evaluated for the reactions of $^{64}$Ni+$^{208}$Pb at
$E_{\rm c.m.}$\,$=$\,268~MeV and $^{58}$Ni+$^{208}$Pb at $E_{\rm c.m.}$\,$=$\,270~MeV
and are compared with available experimental data.

\item[Methods]
Three-dimensional time-dependent Hartree-Fock (TDHF) calculations are carried out for a range
of initial orbital angular momenta with Skyrme SLy4d functional. Quantal diffusion equations,
derived based on the SMF theory, for variances and covariance of neutron and proton numbers
of reaction products are solved, with microscopic drift and diffusion coefficients obtained from
time evolution of occupied single-particle orbitals in TDHF. Secondary deexcitation processes,
both particle evaporation and fission, are simulated by a statistical compound-nucleus deexcitation
model, \texttt{GEMINI++}.

\item[Results]
Dynamics of a fast isospin equilibration process followed by a slow drift toward the mass
symmetry are commonly observed, as expected. Various reaction outcomes are evaluated,
including average mass and charge numbers of reaction products, total kinetic energy loss (TKEL),
scattering angle, contact time, and production cross sections for primary and secondary products.
By comparing with the experimental data, we find that SMF and TDHF quantitatively reproduce
experimental data for few-nucleon-transfer channels around the average values. In contrast,
for many-nucleon-transfer channels, we find that the SMF approach provides much better
description of the experimentally measured isotopic distributions. The results underline the
importance of beyond mean-field effects, especially one-body (mean-field) fluctuations and
correlations, in describing multinucleon transfer processes. Moreover, through a combined
analysis of SMF with a statistical model, \texttt{GEMINI++}, we find a significant contribution
of transfer-induced fission, which is consistent with the experimental observation. In some cases,
the SMF approach overestimates the isotopic width, requiring further improvements of the
theoretical description. Possible ways to improve the description are discussed.

\item[Conclusions]
The SMF approach is designed to describe the quantum many-body problem according to an
ensemble of mean-field trajectories, taking into account part of many-body correlations
in the description. As it requires feasible computational costs comparable to the ordinary TDHF
approach, together with further model improvements, it will be a promising tool in the search for
optimal reaction conditions to produce yet-unknown neutron-rich heavy nuclei through the
multinucleon transfer reaction.
\end{description}
\end{abstract}


\maketitle

\section{INTRODUCTION}

This paper aims to develop a predictive microscopic approach for multinucleon transfer
reactions to guide future experiments for production of yet-unknown neutron-rich heavy
nuclei. The production of unknown neutron-rich isotopes is essential to develop our
understanding of physics of atomic nuclei. Fragmentation, fission, and fusion
processes have been successful to certainly extend the nuclear map of known isotopes
\cite{Thoennessen}. However, there are regions where those methods have difficulty
in producing unstable nuclei, typically located in the north-eastern part of the nuclear landscape,
which is referred to as a `blank spot' \cite{Zagrebaev(2008)}. Study of such neutron-rich
heavy isotopes is of paramount importance not only for nuclear structure, but also for nuclear
astrophysics aspects. Experimental and theoretical investigations of nuclear shape and shell
evolution in the region of neutron-rich nuclei \cite{Otsuka(2020)} and the predicted island
of stability in the superheavy region \cite{Hofmann(2000),Oganessian(2017)} will drive our
understanding of the physics of atomic nuclei. In particular, properties of neutron-rich nuclei
along the neutron magic number $N$\,$=$\,126 are crucial to uncover the detailed pathways
of the $r$-process nucleosynthesis \cite{Kajino(2019)}. Those nuclei correspond to the
last waiting point in the $r$ process, providing the third peak structure at $A\approx$\,195
in the solar abundance. Besides, the region of superheavy nuclei has been explored
by fusion reactions that could produce neutron deficient isotopes with respect to the
$\beta$-stability line \cite{Hofmann(2000),Oganessian(2017)}. The multinucleon
transfer reactions might be a possible alternative to produce neutron-rich superheavy
nuclei in the yet-unreached island of stability, although further investigations are mandatory
\cite{Adamian(2020)}. For the history, current status, and future prospect of the experimental
endeavor for new isotopes production with the multinucleon transfer reaction, we refer
readers to a recent comprehensive review \cite{Adamian(2020)} and references therein.

For the study of multinucleon transfer reactions, semiclassical models such as GRAZING
\cite{GRAZING1,GRAZING2} and complex Wentzel-Kramers-Brillouin (CWKB) \cite{CWKB} have been successfully
used to describe transfer processes at peripheral collisions. GRAZING has been incorporated
with a statistical model to take into account the effect of fission, such as GRAZING-F \cite{Yanez(2015)}
or GRAZING plus GEMINI++ \cite{Wen(2019)}. Although it offers quantitative predictions
for few-nucleon transfers, it substantially underestimates many-nucleon transfer processes
due to missing contributions from deep-inelastic collisions at small impact parameters. The so-called
dinuclear system (DNS) model, initially developed for fusion reactions, has been applied also for
multinucleon transfer and quasifission processes in damped collisions. In the latter model, the
probability distribution for production of various isotopes are derived either by solving a master
equation for mass and/or charge asymmetry \cite{Zhu(2019),Zhang(2019),Guo(2019),Chen(2020)}
or using a simplified statistical expression \cite{Penionzhkevich(2005),Penionzhkevich(2006),Mun(2014),
Mun(2015),Mun(2019)}. To include contributions from peripheral collisions, which are absent
in the DNS model, a simple hybrid called DNS+GRAZING has been considered in the literature
\cite{Welsh(2017),Wen(2017)}. A Langevin-type dynamical model has also been successful in
describing multinucleon transfer, quasifission, and fusion, in a unified way \cite{Zagrebaev(2007)1,
Zagrebaev(2007)2} and has been further improved in Refs.~\cite{Karpov(2017),Saiko(2018)}.
On the other hand, there are microscopic models that treat explicitly nucleonic degrees of freedom
such as improved quantum molecular dynamics (ImQMD) model \cite{Wang(2016),Li(2016),
Zhao(2016),Li(2019)}. Although the latter model still neglects the spin-orbit interaction, it has shown
successes in describing mass, charge, and total kinetic energy (TKE) distributions. While the
above-mentioned approaches have been extensively developed and successfully applied, they rely on
phenomenology to a certain extent . In the present paper, we employ microscopic time-dependent
self-consistent mean-field theories, such as time-dependent Hartree-Fock (TDHF) and its extension,
which contains no adjustable parameters and with no artificial restrictions on the reaction dynamics.
(See, the above mentioned references and review papers \cite{Adamian(2020),Corradi(2009),
Zhang(2018),Sekizawa(2019)}, and references cited therein, for many other applications and discussions.)

The TDHF approach can properly describe the most probable dynamical path in low-energy
heavy-ion reactions, resulting in a good description of total kinetic energy loss (TKEL) and
scattering angle, as well as average neutron and proton numbers. This is supported by an
extended variational principle of Balian and V\'en\'eroni \cite{BV(1981)}, which derives
TDHF as a variation optimized for one-body observables \cite{Simenel(review)}. With the
help of the particle-number projection method \cite{Simenel(2010)}, one can also extract
the probability for production of each isotope. The particle-number projection method has
been used to study multinucleon transfer processes in
$^{16}$O+$^{208}$Pb \cite{Simenel(2010)},
$^{40,48}$Ca+$^{124}$Sn, $^{40}$Ca+$^{208}$Pb, $^{58}$Ni+$^{208}$Pb \cite{KS_KY_MNT,KS_GEMINI},
$^{24}$O+$^{16}$O \cite{KS_KY_PNP},
$^{18}$O+$^{206}$Pb \cite{Bidyut(2015)},
$^{64}$Ni+$^{238}$U \cite{KS_GEMINI,KS_KY_Ni-U},
$^{136}$Xe+$^{198}$Pt \cite{KS_GEMINI},
$^{238}$U+$^{124}$Sn \cite{KS_U-Sn},
$^{16}$O+$^{27}$Al \cite{Bidyut(2017)},
$^{58}$Ni+$^{124}$Sn \cite{Wu(2019)},
$^{132}$Sn+$^{208}$Pb \cite{Jiang(2018)},
$^{136}$Xe,\,$^{132}$Sn+$^{208}$Pb \cite{Jiang(2020)}, and
$^{136}$Xe+$^{194}$Ir \cite{Jiang(2020)2},
at energies around the Coulomb barrier. It has been shown that TDHF works quite well
in describing production cross sections quantitatively for transfer of a few nucleons around
the average values. Recently, the TDHF approach has been combined with a statistical model
to evaluate effects of secondary processes of excited reaction products \cite{KS_GEMINI,Wu(2019),
Jiang(2018),Jiang(2020),Jiang(2020)2,Umar(2017),Guo(2018)2}. It has been quantified that TDHF
underestimates production cross sections for many-nucleon transfer channels. This is related to
the well-known drawback of the TDHF approach, that is, it cannot describe the fluctuations of
the collective dynamics and severely underestimates the fragment mass and charge dispersions
\cite{Koonin(1977),Dasso(1979),Broomfield(2009),Simenel(2011),Williams(2018)}.
One should note that the particle-number projection method does not go beyond TDHF,
but it is just a technique to extract transfer probabilities from the TDHF wave function
after collision \cite{Simenel(2010),KS_KY_MNT}. For a reliable, quantitative description
of processes far apart from the mean trajectory, one must go beyond the standard
TDHF description.

Recently, it has been shown that the description can be improved significantly by the use
of time-dependent random phase approximation (TDRPA) which can be derived from the
extended variational principle of Balian and V\'en\'eroni \cite{BV(1981)}. The variation
suitable for describing dispersions of one-body observables with a single Slater determinant
gives rise to the TDRPA formula that takes into account one-body fluctuations and correlations
around the TDHF average trajectory. TDRPA was applied to deep-inelastic collisions of
$^{16}$O+$^{16}$O \cite{Broomfield(2009)} and $^{40}$Ca+$^{40}$Ca \cite{Simenel(2011)},
showing substantial improvements of the description. Recently, the TDRPA results of the
width of fragment mass distribution for deep-inelastic $^{60}$Ni+$^{60}$Ni collisions
were compared with the experimental data of $^{58}$Ni+$^{60}$Ni at the same
center-of-mass energies, showing a remarkable quantitative agreement \cite{Williams(2018)}.
The results indicate that the one-body fluctuations incorporated by TDRPA are the predominant
mechanism for the mass-width evolution in heavy-ion reactions at low energies. Furthermore,
$^{176}$Yb+$^{176}$Yb collisions were investigated within the TDHF and TDRPA
approaches and primary production cross sections were computed that suggest possible
production of neutron-rich nuclei \cite{Godbey(2019)}. One should note, however, that,
as was shown in Ref.~\cite{Williams(2018)}, the TDRPA formula in the current form cannot
be applied to asymmetric systems, which prevents systematic investigations for various
projectile-target combinations within the TDRPA approach.

In the present paper, we investigate an alternative approach, called the stochastic mean-field
(SMF) approach, proposed by Ayik in 2008 \cite{Ayik(2008)1}, which incorporates beyond
mean-field fluctuations and correlations into the description. The original idea was to model
the quantum many-body problem by an ensemble of time-dependent mean-field solutions
by introducing initial mean-field fluctuations, akin to, in some sense, the derivation of
quantum mechanics from Brownian particles \cite{Nelson(1966)}. Later, it was shown that
the SMF treatment includes more than just the one-body fluctuations and correlations through
a simplified Bogoliubov-Born-Green-Kirkwood-Yvon (BBGKY) hierarchy \cite{Lacroix(2016)}.
It can be shown that the SMF approach, while being applicable also to asymmetric systems,
coincides analytically with the TDRPA formula in the small fluctuation limit \cite{Ayik(2008)1,
Lacroix(2014)1}. In recent years, there have been rapid developments and improvements
in the description. In the initial stage of applications, a semiclassical treatment with the
Wigner transformation and with the Markov approximation was employed \cite{Ayik(2009),
Washiyama(2009),Yilmaz(2011),Yilmaz(2014)1,Ayik(2015)2}. In Ref.~\cite{Ayik(2015)1},
a quantal expression of the diffusion coefficient was proposed, further refined by eliminating
particle states from the expression with the completeness relation \cite{Ayik(2016)}, which
is expressed in terms of the single-particle orbitals from the mean-field theory. The quantal
expression was applied for central collisions of symmetric systems, $^{28}$O+$^{28}$O,
$^{40,48}$Ca+$^{40,48}$Ca, and $^{56}$Ni+$^{56}$Ni, just below the Coulomb barrier,
and also for head-on collisions of $^{238}$U+$^{238}$U \cite{Ayik(2017)}. Finally, the
quantal diffusion description was generalized for non-central collisions \cite{Ayik(2018)}.
The approach was successfully applied to
$^{48}$Ca+$^{238}$U \cite{Ayik(2018)},
$^{58,60}$Ni+$^{60}$Ni \cite{Yilmaz(2018)},
$^{136}$Xe+$^{208}$Pb \cite{Ayik(2019)1,Ayik(2019)2}, and
$^{48}$Ca+$^{208}$Pb \cite{Yilmaz(2020)} systems.
We mention here that the SMF approach has also been applied in other contexts such as
spinordal instabilities of nuclear matter \cite{Ayik(2008)2,Ayik(2009)2,Ayik(2011),O-Yilmaz(2011),
O-Yilmaz(2013),Yilmaz(2015),Acar(2015)}, symmetry breaking \cite{Lacroix(2012)}, Fermionic
Hubbard clusters \cite{Lacroix(2014)2}, as well as nuclear fission \cite{Tanimura(2018)}.
(For other mean-field approaches with stochastic extensions, see discussions in, e.g.,
Refs.~\cite{Lacroix(2014)1,StochasticTDDFT}, and references therein.)

In this work, we analyze the multinucleon transfer processes in the
$^{64}$Ni+$^{208}$Pb reaction at $E_{\rm c.m.}$\,$=$\,268~MeV and the
$^{58}$Ni+$^{208}$Pb reaction at $E_{\rm c.m.}$\,$=$\,270~MeV, for which experimental
data are available \cite{Krolas(2003),Krolas(2010)}. Because of the relatively large isospin
asymmetry of the systems, a fast isospin equilibration process takes place. Also, since the
systems have a relatively large charge product, $Z_{\rm P}Z_{\rm T}$\,$=$\,2296, an onset
of quasifission emerges accompanying a transfer of many nucleons from heavy to light nuclei
that drives the system toward the mass symmetry. The comparison of these two systems
at almost the same center-of-mass energy will reveal detailed reaction mechanisms,
especially isospin dependence of the dynamics. In the experiments by Kr\'olas \textit{et al.}
\cite{Krolas(2003),Krolas(2010)}, a thick target was utilized and a full set of reaction
products were thoroughly analyzed, which were stopped in the target material. They
performed elaborated analyses of in-beam and off-line $\gamma$-$\gamma$ coincidences
\cite{Krolas(2010)}, supplemented with off-line radioactivity measurements \cite{Krolas(2003)}.
Production yields were then identified for abundant isotopes, both projectile-like fragments
(PLFs) and target-like fragments (TLFs), automatically covering the whole angular range,
and from various origins, not only deep-inelastic collisions but also fragments of transfer-induced
fission. The comparison between the measurements and the calculations thus sheds light
on the applicability of theoretical approaches. It is shown clearly that the SMF approach
provides much better description for many-nucleon transfer processes, where TDHF fails
to describe magnitude of production cross sections by orders of magnitude. Finally, the
possibility to produce neutron-rich nuclei along the neutron magic number $N$\,$=$\,126 is discussed.

The article is organized as follows. In Sec.~\ref{sec:method}, the theoretical frameworks
of TDHF and the quantal diffusion approach for multinucleon transfer processes based on
the SMF theory are recalled. In Sec.~\ref{sec:results}, we present the results of TDHF and
SMF calculations for the $^{58,64}$Ni+$^{208}$Pb reactions, which are compared with
the available experimental data. Conclusions are given in Sec.~\ref{sec:conclusions}.

\section{METHODS}\label{sec:method}

\subsection{The TDHF theory}

The TDHF theory in nuclear physics has a long history since the 1970s \cite{Negele(review)}.
With the continuous development of computational technology, it has become a standard tool
to investigate various nuclear dynamics microscopically within the self-consistent mean-field
picture. It is nowadays regarded as a time-dependent energy density functional (TDEDF) approach
rooted with the concept of nuclear density functional theory (DFT) and its time-dependent extension
(TDDFT) \cite{Nakatsukasa(review)}. With the use of a local EDF, the TDHF equation has a generic form:
\begin{equation}
i\hbar\frac{\partial\phi_h^{q}(\bs{r}\sigma,t)}{\partial t}
= \sum_{\sigma'}\hat{h}_{\sigma\sigma'}^{q}(\bs{r},t)\phi_h^{q}(\bs{r}\sigma',t),
\end{equation}
where $\phi_h^{q}(\bs{r}\sigma,t)$ are the $h\,$th occupied (hole) state with spatial, spin,
and isospin coordinates, $\bs{r}$, $\sigma$, and $q$ ($q$\,$=$\,$n$ for neutrons and $q$\,$=$\,$p$
for protons), respectively. $\hat{h}_{\sigma\sigma'}^q(\bs{r},t)$ denotes the single-particle
Hamiltonian which depends on various densities. For instance, the number and the current
densities are expressed in terms of the single-particle orbitals as follows:
\begin{eqnarray}
\rho_q(\bs{r},t) &=& \sum_{h,\sigma}^{\rm occ.}\bigl|\phi_h^{q}(\bs{r}\sigma,t)\bigr|^2,\\
\bs{j}_q(\bs{r},t) &=& \frac{\hbar}{m}\sum_{h,\sigma}^{\rm occ.}{\rm Im}\bigl[\phi_h^{q*}(\bs{r}\sigma,t)\bs{\nabla}\phi_h^{q}(\bs{r}\sigma,t)\bigr].\label{Eq:j_q}
\end{eqnarray}
EDF is constructed so as to reproduce static properties of finite nuclei over a wide range
in the nuclear chart and basic nuclear matter properties in the spirit of nuclear DFT. With
the use of the same form of EDF for static and dynamic calculations (disregarding possible
memory effects in the functional), the TDHF approach offers a unified description of nuclear
structure and dynamics without empirical parameters.

We note that the Pauli exclusion principle is automatically ensured for all times. With the spin-orbit
interaction, the shell effects and deformation, both static and dynamic, are automatically described
in a unified way. It is therefore possible to self-consistently describe complex reaction dynamics
not only nucleon transfer, but also shape deformation (neck formation), surface vibrations,
single-particle excitations, energy and angular momentum dissipation, microscopically from
nucleonic degrees of freedom. The application of the TDHF approach ranges from collective
excitation modes of an isolated nucleus \cite{NakatsukasaYabana(2005),UO(liner-responce),
Maruhn(GDR),Ebata(2010),Fracasso(2012),Pardi(2013),Scamps(2013)LR,Scamps(2014)LR,
Ebata(2014)} to nuclear reactions like transfer \cite{Simenel(2010),KS_KY_MNT,KS_GEMINI,
KS_KY_PNP,Bidyut(2015),KS_KY_Ni-U,KS_U-Sn,Bidyut(2017),Wu(2019),Jiang(2018),
Jiang(2020),Jiang(2020)2,Umar(2008),Iwata(2010),Evers(2011),Umar(2017),KS_SH_Kazimierz},
quasifission \cite{Kedziora(2010),Simenel(2012),Wakhle(2014),Oberacker(2014),Umar(2015)1,
Washiyama(2015),Umar(2015)3,Umar(2016)1,Yu(2017),Guo(2018)2,Zheng(2018)}, fusion
\cite{Keser(2012),Simenel(2013)1,Reinhard(2016),Godbey(2017),Guo(2018)3,Simenel(2007),
Bourgin(2016),Vo-Phuoc(2016)}, and fission \cite{Simenel(2014),Scamps(2015),
Tanimura(2015),Goddard(2015),Goddard(2016),Scamps(2018)2}. For more details
of the TDHF approach and its various applications, see, e.g., Refs.~\cite{Negele(review),
Simenel(review),Nakatsukasa(PTEP),Nakatsukasa(review),TDHF-review(2018),Stevenson(2019),
Sekizawa(2019)}. In this way, the TDHF approach is a versatile tool for studying quantum
many-body dynamics in nuclear systems at low energies. However, the inherent
suppression of fluctuations by the common mean-field needs to be overcome.
The aim of the present paper is to tackle this problem by the SMF approach.

\subsection{The SMF theory}

We recall the basic concepts inherent in the SMF approach. We will omit here
the spin and isospin indexes for simplicity. For a detailed derivation and discussions,
we refer the readers to, e.g., Refs.~\cite{Ayik(2008)1,Lacroix(2014)1,Ayik(2018)}.

In low-energy heavy-ion reactions at energies around the Coulomb barrier, two-body
dissipation would play a minor role owing to the Pauli exclusion principle, and one-body
dissipation presumably plays a predominant role. The observed agreements between
recent TDHF calculations and experimental data offer strong support on this picture
(see, e.g., Refs.~\cite{Williams(2018),KS_KY_Ni-U,Wakhle(2014)}). It is therefore
reasonable to assume that one-body (mean-field) fluctuations, the counterpart of the
one-body dissipation, are the major source for generating a distribution of observables
in nuclear reactions at low energies. Generally, the ground-state wave function of an
atomic nucleus is not a mere single Slater determinant, but rather a superposition of
many Slater determinants, as shown in, e.g., the success of the generator coordinate
method (GCM) for nuclear structure calculations \cite{Ring-Schuck,Egido(2016)},
that can be viewed as quantal zero-point fluctuations of the mean-field potential.

To take into account the mean-field fluctuations, Ayik proposed \cite{Ayik(2008)1}
to introduce fluctuations in the density matrix at the initial time,
\begin{equation}
\rho^\lambda(\bs{r},\bs{r}',t_0) = \sum_{i,j}\phi_i^*(\bs{r},t_0)\rho_{ij}^\lambda\phi_j(\bs{r}',t_0),
\label{Eq:rho_lambda}
\end{equation}
where $\lambda$ labels each stochastically-generated event. Note that the stochastic
elements $\rho_{ij}^\lambda$ in the right hand side of Eq.~(\ref{Eq:rho_lambda})
do not depend on time. The generated density matrices evolve in time independently
from each other according to its own self-consistent mean-field potential, i.e.:
\begin{equation}
i\hbar\frac{d\rho^\lambda}{dt}=\bigl[h[\rho^\lambda(t)],\rho^\lambda\bigr].
\end{equation}
Note that in the SMF approach the stochasticity is introduced only at the initial time,
and the time evolution of the mean-field in each event $\lambda$ itself is not
a stochastic process. The key question, and this is the most important element
behind the SMF theory, is how to imprint the initial fluctuations.

The initial fluctuations are introduced in the following way. Each event $\lambda$
generates the expectation value of a one-body observable, $\bigl<A\bigr>^\lambda$\,$=$\,${\rm Tr}
[\rho^\lambda A]$. In the SMF approach, the original quantum mechanical framework
is then replaced with a statistical treatment. Namely, the expectation value and the
variance of a one-body observable are, respectively, evaluated as \cite{Ulgen(2019)}
\begin{eqnarray}
\overline{\bigl<A\bigr>^\lambda}
&=& {\rm Tr}[\overline{\rho^\lambda}A] = \sum_{ij}\overline{\rho_{ij}^\lambda}A_{ji},
\label{Eq:mean_SMF}\\[-1mm]
\overline{\bigl(\bigl<A\bigr>^\lambda - \overline{\bigl<A\bigr>^\lambda}\bigr)^2}
&=& \overline{\bigl({\rm Tr}[\delta\rho^\lambda A]\bigr)^2}
= \sum_{ijkl}\overline{\delta\rho_{ij}^\lambda\delta\rho_{kl}^\lambda}A_{ji}A_{lk},
\nonumber\\[-3mm]\label{Eq:variance_SMF}
\end{eqnarray}
where $\delta\rho^\lambda$ is the fluctuating part of the density matrix, i.e. $\delta\rho^\lambda=\rho^\lambda-\overline{\rho^\lambda}$. Here and henceforth,
the bar over quantities represents the ensemble average over the stochastically generated
events. On the other hand, for the natural basis satisfying $\bigl<i\big|\rho\big|j\bigr>
=n_i\delta_{ij}$ at the initial time, where $n_i$ are average occupation numbers of
the single-particle states, the quantum mechanical expressions of the expectation value
and the variance of a one-body observable are, respectively, given by \cite{Ulgen(2019)}
\begin{eqnarray}
\bigl<A\bigr> &=& \sum_i n_iA_{ii},
\label{Eq:mean_QM}\\[1mm]
\bigl<A^2\bigr> - \bigl<A\bigr>^2 &=& \frac{1}{2}\sum_{ij}[n_i(1-n_j)+n_j(1-n_i)]A_{ji}A_{ij}.
\nonumber\\[-3mm]\label{Eq:variance_QM}
\end{eqnarray}
The essential point of the SMF theory is that it is designed in such a way that the expectation
value and the variance obtained with the statistical treatment, Eqs.~(\ref{Eq:mean_SMF})
and (\ref{Eq:variance_SMF}), coincide with the quantum expressions, Eqs.~(\ref{Eq:mean_QM})
and (\ref{Eq:variance_QM}), respectively, at the initial time. It is accomplished by setting
the initial fluctuations according to \cite{Ayik(2008)1}
\begin{eqnarray}
\overline{\rho_{ij}^\lambda} &=& n_i\delta_{ij}, \label{Eq:SMF1}\\
\overline{\delta\rho_{ij}^\lambda\delta\rho_{kl}^{\lambda}}
&=& \frac{1}{2}[n_i(1-n_j) + n_j(1-n_i)]\delta_{kj}\delta_{li}. \label{Eq:SMF2}
\end{eqnarray}
Since the fluctuating components of the density matrix have zero mean, by construction,
an ensemble average of those events reproduces the ordinary mean-field (TDHF) result.

We note that when the initial state has zero temperature such as the ground state of
projectile and target nuclei before collision, the average occupation numbers $n_i$ are
zero or one. If an observable is diagonal at the initial state, $A_{ij}=A_i\delta_{ij}$,
such as particle number operators for the projectile and target, as seen in Eq.~(\ref{Eq:variance_QM}),
the variance of such observables are strictly zero and therefore they do not exhibit
fluctuations at the initial state. If the initial state has a finite temperature, in the
case of induced fission of a compound nucleus for instance, the average values
of the occupation numbers are determined by the Fermi-Dirac distribution.

As mentioned in the introduction, it is worth noting here that, although the SMF approach
was originally proposed to take into account one-body (mean-field) fluctuations at the
initial time, it has been shown that it grasps part of many-body correlations through
a simplified BBGKY hierarchy \cite{Lacroix(2016)}. In addition, in the original formulation
of the SMF approach \cite{Ayik(2008)1}, the stochastic matrix elements $\delta\rho_{ij}^\lambda$
are assumed to be uncorrelated Gaussian random numbers with zero mean. Recently,
in Ref.~\cite{Ulgen(2019)}, by analyzing higher-order moments of one-body observables
(the first and the second moments correspond to the mean and the variance, respectively),
it has been shown that the description can be further improved by relaxing the Gaussian
assumption (see also  Ref.~\cite{Yilmaz(2014)2}). In the present article, however,
we adopt the Gaussian assumption for the stochastic matrix elements, which allow
us to formulate a quantal diffusion description for multinucleon exchanges, as described
in Sec.~\ref{Sec:diffusion}. We leave further improvements of such model ingredients
as future works.

\subsection{The quantal diffusion description}\label{Sec:diffusion}

\begin{figure} [t]
\includegraphics[width=8.6cm]{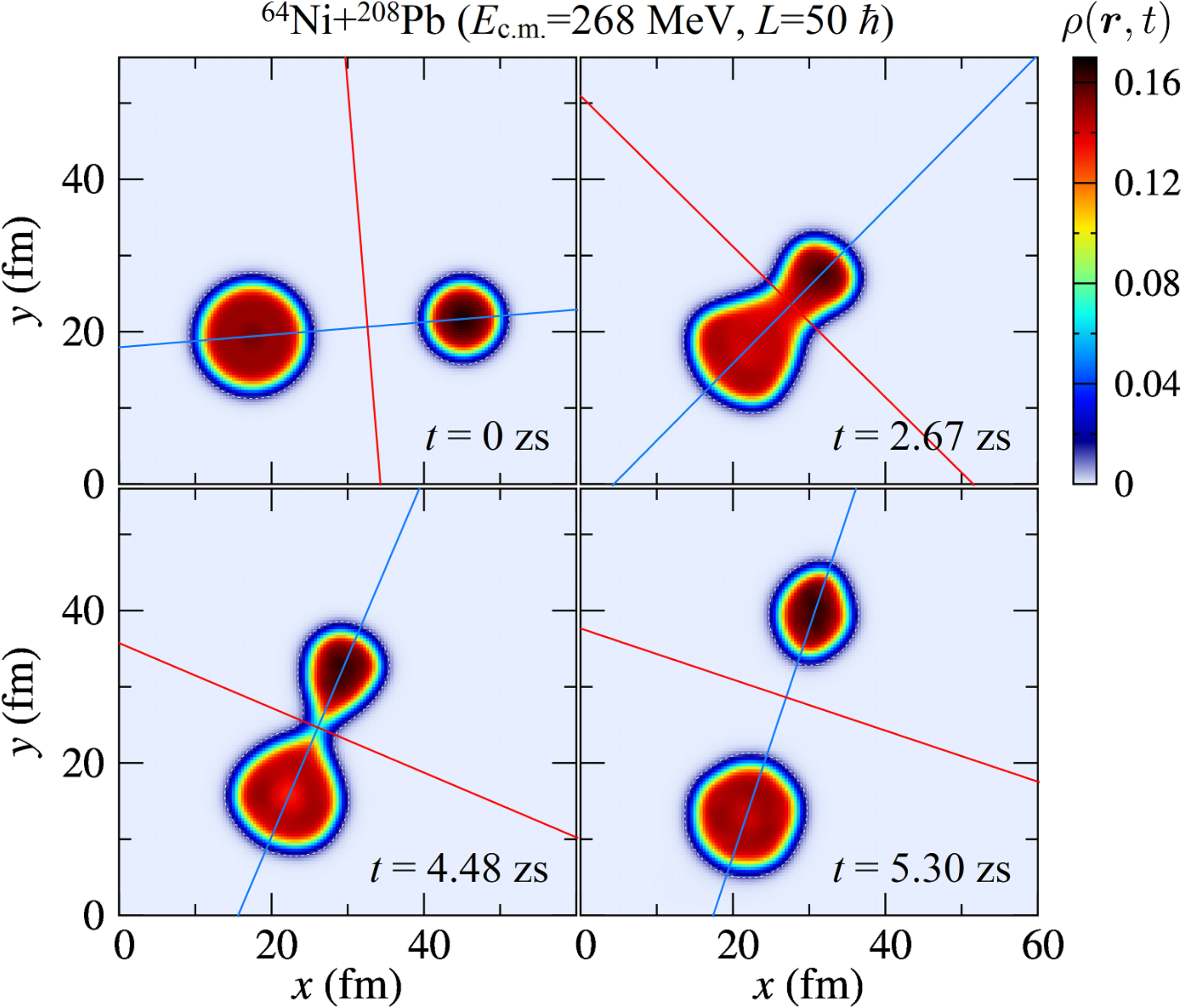}\vspace{-1mm}
\caption{
Snapshots of density distribution in the $^{64}$Ni+$^{208}$Pb reaction at $E_{\rm c.m.}=268$~MeV
with $L$\,$=$\,50$\hbar$ are shown in the reaction plane. The blue line indicates the elongation
axis which passes through the centers of mass of the projectile-like and target-like subsystems. The
red line indicates the position of the window plane. A contour of $\rho$\,$=$\,0.01~fm$^{-3}$ is indicated
by dotted lines. Elapsed time is indicated in each panel in zeptoseconds (1\,zs\,$=$\,10$^{-21}$\,s).
}\vspace{-3mm}
\label{Fig:rho(t)}
\end{figure}

When dinuclear structure is maintained during a collision (cf.~Fig.~\ref{Fig:rho(t)}),
it is possible to define a window at the neck region and derive quantal diffusion descriptions
for multinucleon exchanges. Namely, it allows us to define neutron and proton numbers
of a projectile-like subsystem, $N_1^\lambda(t)$ and $Z_1^\lambda(t)$, respectively,
as macroscopic variables. Then, the nucleon exchange can be described as a diffusion
process \cite{Randrup(1979)}. The evolution of the neutron and proton numbers is
described by the Langevin equation:
\begin{equation}
\frac{d}{dt}\begin{pmatrix}N_1^\lambda(t)\\ Z_1^\lambda(t)\end{pmatrix}=
\int g(x')\begin{pmatrix}\hat{\bs{e}}\bs{\cdot}\bs{j}_n^\lambda(\bs{r},t)\\ \hat{\bs{e}}\bs{\cdot}\bs{j}_p^\lambda(\bs{r},t)\end{pmatrix}d\bs{r}=
\begin{pmatrix}\nu_n^\lambda(t)\\ \nu_p^\lambda(t)\end{pmatrix},
\label{Eq:Langevin}
\end{equation}
where $\bs{j}_q^\lambda(t)$ and $\nu_q^\lambda(t)$ ($q$\,$=$\,$n$ or $p$)
are the current densities and drift coefficients in the event $\lambda$, respectively.
The unit vector $\hat{\bs{e}}$ is perpendicular to the window plane and directed
along the relative position vector from the center of the target-like subsystem to
the center of the projectile-like subsystem, $\hat{\bs{e}}(t)=\cos\theta(t)\hat
{\bs{x}}+\sin\theta(t)\hat{\bs{y}}$. Here, $\theta(t)$ represents the (initially)
smaller angle between the elongation axis of the colliding system and the collision axis.
The elongation axis and the rotation angle $\theta(t)$ can be determined by
diagonalizing the mass quadrupole matrix as described in Refs.~\cite{Yilmaz(2014)1,
Ayik(2018)}. The smoothing function, $g(x')=\frac{1}{\sqrt{2\pi}\kappa}
\exp[-x^{\prime2}/2\kappa^2]$, extracts the contribution at the vicinity of
the window plane, where $x'=\hat{\bs{e}}\bs{\cdot}(\bs{r}-\bs{r}_0)$
with $\bs{r}_0$ indicating the center of the window plane. The smoothing
parameter $\kappa$\,$=$\,1.0\,fm is used, which is the same order of the
lattice spacing, as described in Ref.~\cite{Ayik(2016)}.

To obtain equations for the variances and the covariance, we use the stochastic
part of the Langevin equation (\ref{Eq:Langevin}), which is linearized assuming
small fluctuations around the mean evolution:
\begin{eqnarray}
\frac{d}{dt}\begin{pmatrix}\delta N_1^\lambda(t)\\ \delta Z_1^\lambda(t)\end{pmatrix}
&=&
\begin{pmatrix}
\frac{\partial\nu_n}{\partial Z_1}\delta Z_1^\lambda(t)
+\frac{\partial\nu_n}{\partial N_1}\delta N_1^\lambda(t)\\
\frac{\partial\nu_p}{\partial Z_1}\delta Z_1^\lambda(t)
+\frac{\partial\nu_p}{\partial N_1}\delta N_1^\lambda(t)
\end{pmatrix}
+
\begin{pmatrix}
\delta\nu_n^\lambda(t)\\
\delta\nu_p^\lambda(t)
\end{pmatrix}.\nonumber\\
\label{Eq:Langevin_2}
\end{eqnarray}
Here, $\delta N_1^\lambda$\,$=$\,$N_1^\lambda-N_1$ and $\delta Z_1^\lambda$\,$=$\,$Z_1^\lambda-Z_1$
denote the stochastic part of neutron and proton numbers of the projectile-like
subsystem, respectively, with $N_1$\,$=$\,$\overline{N_1^\lambda}$ and
$Z_1$\,$=$\,$\overline{Z_1^\lambda}$. Similarly,
$\delta\nu_n^\lambda$\,$=$\,$\nu_n^\lambda-\nu_n$ and
$\delta\nu_p^\lambda$\,$=$\,$\nu_p^\lambda-\nu_p$ denote the
stochastic parts of neutron and proton drift coefficients, respectively, with
$\nu_n$\,$=$\,$\overline{\nu_n^\lambda}$ and $\nu_p$\,$=$\,$\overline{\nu_p^\lambda}$.
The derivatives of the drift coefficients are evaluated at the mean trajectory.
Multiplying both sides of Eq.~(\ref{Eq:Langevin_2}) by $\delta N_1^\lambda$
and $\delta Z_1^\lambda$ and taking the ensemble average, one can derive
a set of coupled partial differential equations \cite{Ayik(2015)2,Ayik(2018)}:
\begin{eqnarray}
\frac{\partial\sigma_{NN}^2}{\partial t} &=&
2\frac{\partial\nu_n}{\partial N_1}\sigma_{NN}^2
+2\frac{\partial\nu_n}{\partial Z_1}\sigma_{NZ}^2
+2D_{NN},
\label{Eq:sigma2_NN}\\
\frac{\partial\sigma_{ZZ}^2}{\partial t} &=&
2\frac{\partial\nu_p}{\partial Z_1}\sigma_{ZZ}^2
+2\frac{\partial\nu_p}{\partial N_1}\sigma_{NZ}^2
+2D_{ZZ},
\label{Eq:sigma2_ZZ}\\
\frac{\partial\sigma_{NZ}^2}{\partial t} &=&
\frac{\partial\nu_p}{\partial N_1}\sigma_{NN}^2
+\frac{\partial\nu_n}{\partial Z_1}\sigma_{ZZ}^2
+\Bigl(\frac{\partial\nu_n}{\partial N_1}+\frac{\partial\nu_p}{\partial Z_1}\Bigr)\sigma_{NZ}^2,\nonumber\\[-1mm]
\label{Eq:sigma2_NZ}
\end{eqnarray}
with the initial conditions $\sigma_{NN}$\,$=$\,$\sigma_{ZZ}$\,$=$\,$\sigma_{NZ}$\,$=$\,0
at $t$\,$=$\,0. Note that the particle number is not fluctuating at the initial time.
$\sigma_{NN}^2$\,$=$\,$\overline{(N_1^\lambda-\overline{N_1^\lambda})^2}$ and
$\sigma_{ZZ}^2$\,$=$\,$\overline{(Z_1^\lambda-\overline{Z_1^\lambda})^2}$
are the variances of neutron and proton numbers, respectively, and
$\sigma_{NZ}^2$\,$=$\,$\overline{(N_1^\lambda-\overline{N_1^\lambda})
(Z_1^\lambda-\overline{Z_1^\lambda})}$ is the covariance (or the mixed variance)
of neutron and proton numbers. $D_{NN}$ and $D_{ZZ}$ are the quantal diffusion
coefficients of neutron and proton exchanges, respectively. The same set of partial differential
equations was employed in phenomenological nucleon exchange models for deep-inelastic
collisions \cite{Schroder(1981),Merchant(1981)}. It is worth emphasizing that all ingredients
of Eqs.~(\ref{Eq:sigma2_NN})--(\ref{Eq:sigma2_NZ}) are determined from the time
evolution of the single-particle orbitals in TDHF, as described in Sec.~\ref{Sec:TransportCoef}.
Therefore, it does not actually require us either to generate an ensemble of mean-field
trajectories or to specify magnitude of fluctuations of $\delta\rho_{ij}^\lambda$.
In practice, only a single TDHF calculation is sufficient to solve Eqs.~(\ref{Eq:sigma2_NN})--(\ref{Eq:sigma2_NZ}),
for a given set of initial conditions. Thus, a systematic investigation is feasible
with moderate computational costs comparable to ordinary TDHF calculations.

\subsection{Transport coefficients}\label{Sec:TransportCoef}

To solve Eqs.~(\ref{Eq:sigma2_NN})--(\ref{Eq:sigma2_NZ}), we need to have the
drift and diffusion coefficients. While the mean drift coefficients, $\nu_n$ and $\nu_p$,
can be evaluated directly from the net mean currents [Eq.~(\ref{Eq:j_q})] passing
through the window in TDHF, it is not straightforward to take derivatives of them
with respect to the neutron and proton numbers of the projectile-like subsystem
[cf. Eqs.~(\ref{Eq:sigma2_NN})--(\ref{Eq:sigma2_NZ})]. We note that the
Langevin equation (\ref{Eq:Langevin_2}) describes the evolution of macroscopic
variables $N_1$ and $Z_1$ in the overdamped limit in which the inertial terms
and the diffusion coefficients of the conjugate momentum of $N_1$ and $Z_1$
do not appear. To evaluate the derivatives of the drift coefficients, we employ
the Einstein relations in the over-damped limit,
\begin{eqnarray}
\nu_n(t) &=& -\frac{D_{NN}(t)}{T^*}\frac{\partial}{\partial N_1}U(N_1,Z_1),\label{Eq:nu_n_Einstein}\\
\nu_p(t) &=& -\frac{D_{ZZ}(t)}{T^*}\frac{\partial}{\partial Z_1}U(N_1,Z_1),\label{Eq:nu_p_Einstein}
\end{eqnarray}
where $T^*$ denotes the effective temperature and $U(N_1,Z_1)$ is the potential
energy surface of the colliding dinuclear system. The quantities $D_{NN}$ and $D_{ZZ}$
are the diffusion coefficients of the macroscopic variables $N_1$ and $Z_1$, not of
the conjugate momentum variables. The effective temperature was introduced by
Randrup in Ref.~\cite{Randrup(1979)}; see also Ref.~\cite{Feldmeier(1987)}.
In calculations of the derivative of drift coefficients, we do not need an explicit expression
of the effective temperature. Only ratios of the curvature parameters and the effective
temperature appear in Eqs.~(\ref{Eq:nu_n_Einstein}) and (\ref{Eq:nu_p_Einstein}).
It is possible to calculate these ratios, which are referred to as the reduced curvature
parameters, $\alpha=a/T^*$ and $\beta=b/T^*$, in terms of the mean drift path
of the collision. The details of determination of the driving potential and the derivatives
of the mean drift coefficients are given in Appendix~\ref{Appendix1}.

The diffusion coefficient is related to the auto-correlation function of the stochastic
part of the drift coefficients, $\delta\nu_q^\lambda$ \cite{Gardiner(1991),Weiss(1999)}:
\begin{equation}
D_{qq}(t) = \int_0^t \overline{\delta\nu_q^\lambda(t)\delta\nu_q^\lambda(t')}dt',
\end{equation}
where
\begin{eqnarray}
\delta\nu_q^\lambda(t)
&=& \hat{\bs{e}}\bs{\cdot}\frac{\hbar}{m}\,\sum_{i,j}\int d\bs{r}\,g(x') \nonumber\\
&\times&\;
{\rm Im}\bigl[\phi_i^{q*}(\bs{r},t;\lambda)\bs{\nabla}\phi_j^{q}(\bs{r},t;\lambda)\bigr]\delta\rho_{ij}^\lambda.
\label{Eq:dnu}
\end{eqnarray}
Note that the summation over $i$ and $j$ in Eq.~(\ref{Eq:dnu}) is taken for
all the complete set of single-particle orbitals including unoccupied (particle) states.

By using the main postulate of the SMF approach, Eq.~(\ref{Eq:SMF2}), together
with the completeness relation in the diabatic approximation, it is possible to eliminate
the unoccupied states from the expression \cite{Ayik(2016)}. As a result, the diffusion
coefficients, including memory (non-Markovian) effects, are determined entirely by
the occupied states in TDHF and are given by \cite{Ayik(2016),Ayik(2018)}
\begin{eqnarray}
D_{qq}(t)
&=& \int_0^t d\tau \int d\bs{r}\;\tilde{g}(x')
\Bigl(
G_{\rm T}(\tau)J_{\rm T}^{q}(\bs{r},t-\tfrac{\tau}{2}) \nonumber\\[-0.5mm]
&&\hspace{24.48mm}+\;G_{\rm P}(\tau)J_{\rm P}^{q}(\bs{r},t-\tfrac{\tau}{2})
\Bigr) \nonumber\\[-0.5mm]
&-&
\int_0^t d\tau\;{\rm Re}
\Bigg[
\sum_{h'\in{\rm P},h\in{\rm T}}^{\rm occ.}A_{h'h}^{q}(t)A_{h'h}^{q*}(t-\tau) \nonumber\\[-0.5mm]
&&\hspace{13.54mm}+\sum_{h'\in{\rm T},h\in{\rm P}}^{\rm occ.}A_{h'h}^{q}(t)A_{h'h}^{q*}(t-\tau)
\Bigg],\hspace{3mm}
\label{Eq:D_qq}
\end{eqnarray}
where $\tilde{g}(x') = \frac{1}{\sqrt{\pi}\kappa'}\exp[-(x'/\kappa')^2]$ is
another smoothing function with a dispersion $\kappa'=0.5$~fm. $G_{\rm P(T)}
(\tau) = \frac{1}{\sqrt{4\pi}\tau_0}\exp[-(\tau/2\tau_0)^2]$ is the averaged
memory kernel for hole states. The memory time is given by $\tau_0$\,$=$\,$\kappa'/|u_0|$,
where $u_0$ stands for the average flow speed of hole states across the window.
$J_\mu^{q}$ denotes the sum of magnitude of the current densities perpendicular
to the window plane, whose contribution comes only from hole states which initially
belong to either projectile ($\mu$\,$=$\,P) or target ($\mu$\,$=$\,T), i.e.,
\begin{eqnarray}
J_\mu^{q}(\bs{r},t)
= \frac{\hbar}{m}\,\sum_{h\in\mu}^{\rm occ.}
\Big| \hat{\bs{e}}\bs{\cdot}{\rm Im}\bigl[\phi_h^{q*}(\bs{r},t)\bs{\nabla}\phi_h^{q}(\bs{r},t)\bigr] \Big|.
\end{eqnarray}
The hole-hole matrix elements, $A_{h'h}^{q}(t)$, are given by
\begin{eqnarray}
&&A_{h'h}^{q}(t) = \hat{\bs{e}}\bs{\cdot}\frac{\hbar}{2m}\int d\bs{r}\;g(x') \nonumber\\
&&\times\,\Bigl(
\phi_{h'}^{q*}(\bs{r},t)\bs{\nabla}\phi_h^{q}(\bs{r},t)
-\phi_{h}^{q*}(\bs{r},t)\bs{\nabla}\phi_{h'}^{q}(\bs{r},t)
\Bigr).\;\;\;\;\;
\end{eqnarray}
(See, Refs.~\cite{Ayik(2016),Ayik(2018)}, for more details.)

The first term in the quantal diffusion coefficient (\ref{Eq:D_qq}) represents the sum
of the nucleon currents between two subsystems across the window, which is integrated
over the memory. It resembles the diffusion coefficient in the random walk problem,
which is given by the sum of the rate for forward and backward steps \cite{Randrup(1979),
Gardiner(1991),Weiss(1999)}. On the other hand, the second term represents the
Pauli blocking effect in nucleon transfer processes, which does not have a classical
counterpart. In this way, the diffusion coefficients, which govern the fluctuation
mechanism of the collective motion, can be determined entirely from the occupied
single-particle orbitals in TDHF. It is rational because the one-body dissipation
mechanism does present within the TDHF approach, which is related to the
fluctuation mechanism as stated in the fluctuation-dissipation theorem.

\subsection{Primary production cross sections}

To evaluate production cross sections, we need to compute the probability for
production of each isotope. Within the TDHF approach, one can employ the
particle-number projection method to obtain the probability to find $n$
nucleons in a reaction product \cite{Simenel(2010),KS_KY_MNT},
\begin{equation}
P_n^{(q)}(b) = \frac{1}{2\pi}\int_0^{2\pi}e^{in\theta}
\det\bigl\{\bigl<\phi_i^q\big|e^{-i\hat{N}_1^{(q)}\theta}\big|\phi_j^q\bigr>\bigr\}\,d\theta,
\end{equation}
where $\hat{N}_1^{(q)}$ denotes the number operator for neutrons
($q$\,$=$\,$n$) or protons ($q$\,$=$\,$p$) in a fragment 1 and $b$ is the
impact parameter. The probability distribution, $P_{N,Z}(b)$, for the
production of nuclei specified by ($N,Z$) has a product form,
\begin{equation}
P_{N,Z}^{\rm TDHF}(b) = P_N^{(n)}(b)P_Z^{(p)}(b).
\label{Eq:P_NZ_TDHF}
\end{equation}
Therefore, there is essentially no correlation between neutron and proton transfers
beyond mean-field (meaning that, e.g., neutrons and protons can still transfer in the
same direction via shape evolution) in the TDHF approach. In other words, the covariance,
$\sigma_{NZ}^2=\bigl<\hat{N}_1^{(n)}\hat{N}_1^{(p)}\bigr>-\bigl<\hat{N}_1^{(n)}\bigr>
\bigl<\hat{N}_1^{(p)}\bigr>$, is strictly zero in TDHF, by construction. As already mentioned
in the introduction, the particle-number projection method is merely a way to extract the
probabilities from the TDHF wave function after collision \cite{Simenel(2010),KS_KY_MNT}.
Thus, it misses correlations between neutron and proton transfers and underestimates
the widths of neutron and proton number distributions.

In the SMF approach, by solving the set of quantal diffusion equations,
Eqs.~(\ref{Eq:sigma2_NN})--(\ref{Eq:sigma2_NZ}), we can obtain the
variances and the covariance of neutron and proton numbers of reaction products.
For uncorrelated Gaussian random numbers, it can be shown that the Langevin equation
(\ref{Eq:Langevin}) is equivalent to the Fokker-Plank description for the probability
distribution $P_{N,Z}(b)$ \cite{Risken(1996)}. In particular when the drift coefficients
have linear dependence as in the present case of the linearized form of the Langevin
equation, the probability distribution is determined by a correlated Gaussian function.
By employing this equivalence, we can write down the probability distribution,
$P_{N,Z}(b)$, as follows:
\begin{equation}
P_{N,Z}^{\rm SMF}(b) = \frac{\exp[-C_{N,Z}(b)]}{2\pi\sigma_{NN}(b)\sigma_{ZZ}(b)\sqrt{1-\eta_b^2}},
\label{Eq:P_NZ_SMF}
\end{equation}
with
\begin{eqnarray}
C_{N,Z}(b)
&=&
\frac{1}{2(1-\eta_b^2)}\biggl[
\biggl(\frac{N-\bar{N}_b}{\sigma_{NN}(b)}\biggr)^2
+\biggl(\frac{Z-\bar{Z}_b}{\sigma_{ZZ}(b)}\biggr)^2 \nonumber\\[1mm]
&&\hspace{11mm}
-2\eta_b\biggl(\frac{N-\bar{N}_b}{\sigma_{NN}(b)}\biggr)\biggl(\frac{Z-\bar{Z}_b}{\sigma_{ZZ}(b)}\biggr)
\biggr],
\end{eqnarray}
where $\eta_b$ is the correlation coefficient, defined by the ratio of the
covariance to the product of neutron and proton dispersions, i.e., $\eta_b\equiv
\sigma_{NZ}^2(b)/\sigma_{NN}(b)\sigma_{ZZ}(b)$. $\bar{N}_b$ and
$\bar{Z}_b$ denote, respectively, the mean neutron and proton numbers
of the reaction product in TDHF.

The production cross sections for \textit{primary} reaction products (i.e.,
just after the collision before deexcitation) are then evaluated by an
integration over the impact parameter:
\begin{equation}
\sigma(N,Z) = 2\pi\int_{b_{\rm min}}^{b_{\rm max}} b\,P_{N,Z}(b)\,db,
\end{equation}
where the minimum and the maximum values of the impact parameters,
$b_{\rm min}$ and $b_{\rm max}$, are chosen according to the angular
coverage of the corresponding experiment. Note that after multinucleon
transfer processes reaction products can be highly excited, and one must
evaluate effects of secondary deexcitation processes to make a direct
comparison with experimental data.

\subsection{Secondary production cross sections}\label{sec:TXE}

For the evaluation of the production cross sections for \textit{secondary}
reaction products (i.e., after deexcitation via particle evaporation,
fission, and $\gamma$-ray emissions), we employ a statistical model for
compound-nucleus disintegration processes. To this end, we follow the
strategy as in Ref.~\cite{KS_GEMINI}. In Ref.~\cite{KS_GEMINI},
the average total excitation energy (the sum of excitation energies of
a PLF and a TLF) was estimated by
\begin{equation}
E_{N,Z}^*(b) = E_{\rm c.m.} - E_{\rm kin}^\infty(b) + Q_{\rm gg}(N,Z),
\end{equation}
where $Q_{\rm gg}(N,Z)$ denotes the ground-to-ground $Q$ value for the
exit channel involving a nucleus specified by $(N,Z)$. Here, $E_{\rm kin}^\infty(b)$
denotes the asymptotic value of TKE of outgoing fragments for the average products
in TDHF. The average total excitation energy is then distributed to PLFs and TLFs
in an appropriate way. This prescription was also used by other authors
\cite{Jiang(2018),Wu(2019),Jiang(2020),Jiang(2020)2}.

However, in the present systems under study as discussed in Sec.~\ref{sec:secondary},
reaction products involve transfer of many (more than 10) protons due to the quantal
diffusion mechanism. In such a case, the Coulomb potential at the scission configuration
can be substantially different from that of the mean trajectory. To have a feeling of it,
imagine that we have touching Ni and Pb nuclei at distance $R$\,$=$\,$1.2(A_{\rm P}^{1/3}
$\,+\,$A_{\rm T}^{1/3})$\,$\simeq$\,12\,fm, for which we have the Coulomb
potential $V_{\rm C}$\,$\simeq$\,275.5\,MeV. If we exchange protons between
those nuclei, keeping the $R$ value unchanged, the Coulomb potential shall be
$V'_{\rm C}$\,$=$\,$(Z_{\rm P}$\,+\,$\Delta Z)(Z_{\rm T}$\,$-$\,$\Delta Z)e^2/R$,
where $\Delta Z$ denotes the number of exchanged protons. The difference,
$\Delta V_{\rm C}$\,$=$\,$V'_{\rm C}$\,$-$\,$V_{\rm C}$, becomes, e.g.,
$+$6.4\,MeV ($-$6.6\,MeV) for $\Delta Z$\,$=$\,$+$1 ($-$1),
$+$12.5\,MeV ($-$13.4\,MeV) for $\Delta Z$\,$=$\,$+$2 ($-$2), $\dots$,
$+$52.8\,MeV ($-$76.8\,MeV) for $\Delta Z$\,$=$\,$+$10 ($-$10), and so on.
The increase (decrease) of the Coulomb potential results in increase (decrease)
of TKE and, thus, the total excitation energy will be decreased (increased).
This is a crude estimation, as the $R$ value could be larger due to fragment
deformation, but it suggests that the secondary cross sections may be affected.

\begin{figure} [t]
\includegraphics[width=6.8cm]{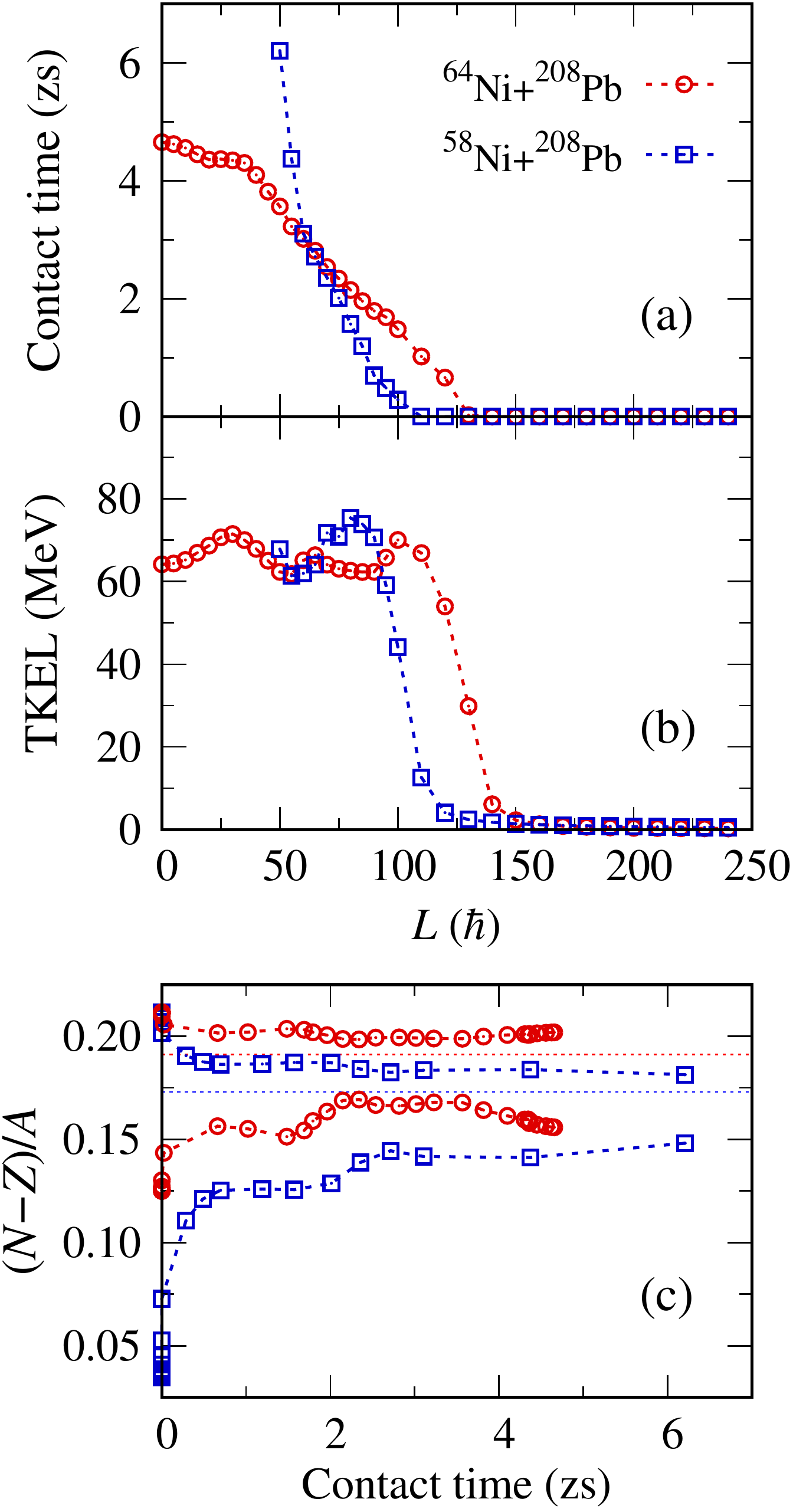}\vspace{-1mm}
\caption{
Results of the TDHF calculations for the $^{64}$Ni+$^{208}$Pb reaction at $E_{\rm c.m.}$\,$=$\,268~MeV
(red open circles) and the $^{58}$Ni+$^{208}$Pb reaction at $E_{\rm c.m.}$\,$=$\,270~MeV
(blue open squares). In panels (a) and (b), contact time and average total kinetic energy loss (TKEL)
are shown, respectively, as a function of the initial orbital angular momentum. In panel (c), average
charge asymmetries, $(N-Z)/A$, for a PLF and a TLF are shown as a function of the contact time.
The contact time is shown in zeptoseconds (1\,zs\,$=$\,10$^{-21}$\,s).
}\vspace{-3mm}
\label{Fig:TKEL}
\end{figure}

Therefore, to grasp the possible energy change due to proton transfers,
we modify the expression of the average total excitation energy as follows:
\begin{eqnarray}
E_{N,Z}^*(b) = E_{\rm c.m.} - E_{\rm kin}^\infty(b) + Q_{\rm gg}(N,Z) - \Delta V_{\rm C}(b,Z).
\nonumber\\[-0.5mm]\label{Eq:E_ex}
\end{eqnarray}
Here, the additional Coulomb correction term is defined as
\begin{equation}
\Delta V_{\rm C}(b,Z) = \frac{Z(Z_{\rm tot}-Z) - \bar{Z}_1(b,t_c)\bar{Z}_2(b,t_c)}{R(b,t_c)}e^2,
\end{equation}
where $Z_{\rm tot}$\,$=$\,$Z_{\rm P}+Z_{\rm T}$ is the total number
of protons in the system. $t_c$ is chosen in the following way: (1) When
two nuclei touch in the course of collision, $t_c$ is taken as the instance
at which a dinuclear system splits, assuming that TKE is determined by the
Coulomb potential at scission; (2) When two nuclei do not touch, $t_c$ is
taken as the instance at the turning point, assuming that the proton transfer
occurs at the closest approach. Processes with a finite contact time [see,
Sec.~\ref{sec:average} and Fig.~\ref{Fig:TKEL}(a)] are regarded as
``touched.'' $\bar{Z}_\mu$ ($\mu$\,$=$\,$1,2$) and $R$ represent the
average number of protons in respective fragments and the relative distance
between them, respectively, at the time $t_c$. In this way, the excitation
energy becomes effectively transfer channel dependent by $Q_{\rm gg}$
and $\Delta V_{\rm C}$. The estimated average total excitation energy
(\ref{Eq:E_ex}) is then distributed to reaction products proportional to
their mass. We should keep in mind that non-equilibrium excitation
energy division may be possible especially for asymmetric systems (see,
e.g., Refs.~\cite{Hilscher(1979),Vandenbosch(1984),Feldmeier(1987)}).
With the average total angular momentum of each reaction product, secondary
processes are simulated by a statistical compound-nucleus deexcitation model,
\texttt{GEMINI++} \cite{GEMINI++,Charity(2010),Mancusi(2010)}, which
includes particle evaporation and fission in competition with $\gamma$-ray emission.

\begin{figure} [t]
\includegraphics[width=\columnwidth]{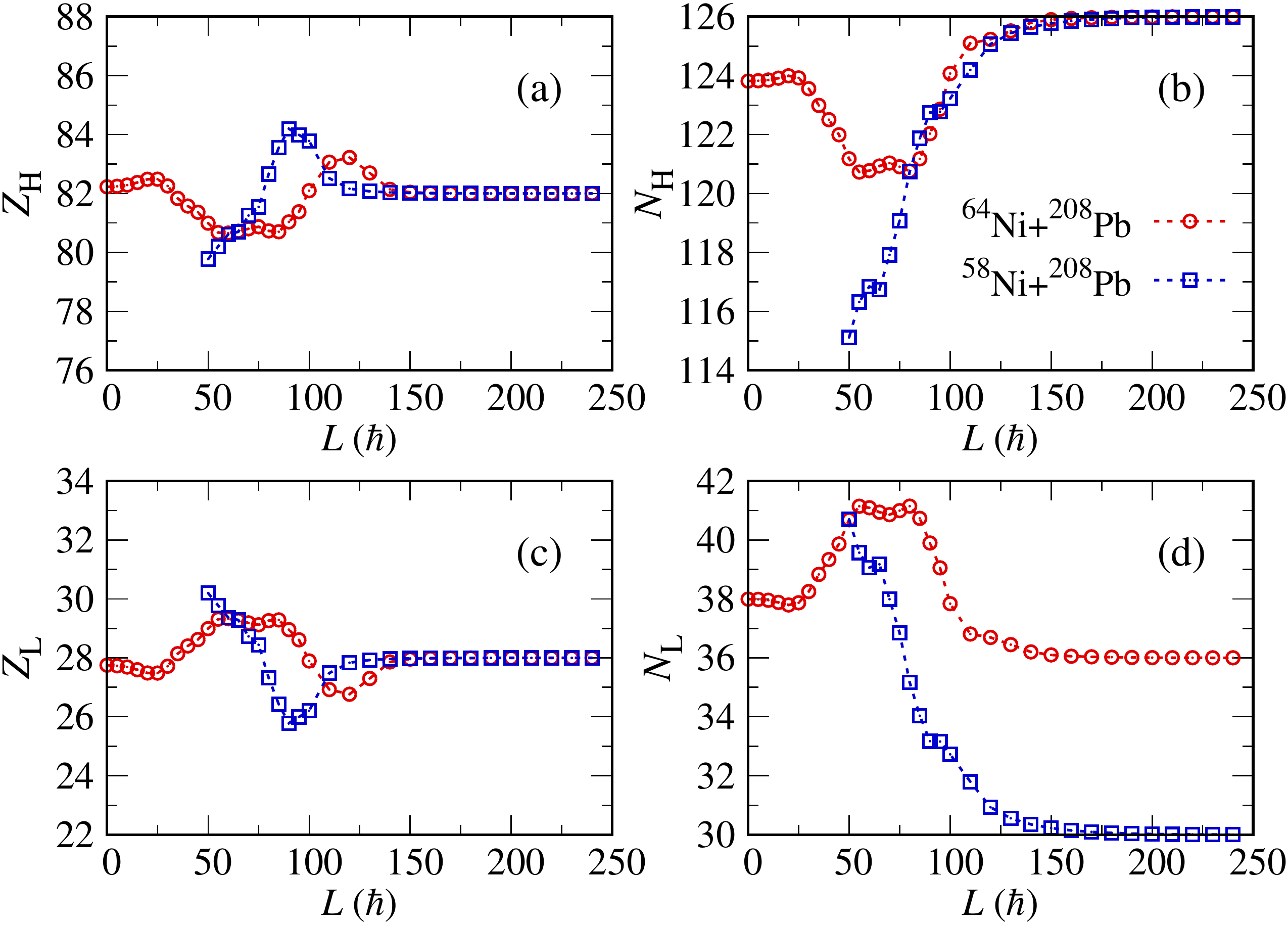}\vspace{-1mm}
\caption{
Results of the TDHF calculations for the $^{64}$Ni+$^{208}$Pb reaction at $E_{\rm c.m.}$\,$=$\,268~MeV
(red open circles) and the $^{58}$Ni+$^{208}$Pb reaction at $E_{\rm c.m.}$\,$=$\,270~MeV
(blue open squares). In panels (a) and (b), average proton and neutron numbers of heavier (target-like)
fragments, $Z_{\rm H}$ and $N_{\rm H}$, are shown, respectively, as a function of the initial
orbital angular momentum. In panels (c) and (d), those for lighter (projectile-like) fragments,
$Z_{\rm L}$ and $N_{\rm H}$, are shown.
}\vspace{-3mm}
\label{Fig:N_and_Z}
\end{figure}

\subsection{Computational details}

To perform the SMF calculations, three-dimensional (3D) parallel TDHF code \cite{KS_KY_MNT}
has been extended and applied. For the EDF, we employ the Skyrme SLy4d functional \cite{Kim(1997)}.
In the code, single-particle orbitals are represented on 3D uniform grid points with the isolated
(box) boundary condition. The grid spacing is set to 0.8\,fm. First and second spatial derivatives
are computed with the 11-point finite-difference formulas. The Coulomb potential is computed by
Fourier transforms. A computational box of $24^3$\,fm$^3$ was used for the ground-state
calculations, while a box of $65\times60\times24$\,fm$^3$ was used for the reaction calculations.
With this setting, we find that the ground state of $^{58}$Ni is of prolate shape with $\beta$\,$=$\,0.12,
while $^{64}$Ni exhibits a shape with $\beta$\,$=$\,0.14 with $\gamma$\,$=$\,47$^\circ$.
$^{208}$Pb is of course of spherical shape. We place those deformed projectiles to have an
orientation with the smallest $Q_{22}$ to be lying in the reaction plane. For the time evolution,
the fourth-order Taylor expansion method was used with a single predictor-corrector step with
$\Delta t$\,$=$\,0.2\,fm/$c$. The initial separation distance between the projectile and target
nuclei was set to 28\,fm along the collision axis. The time evolution was stopped when the relative
distance between PLF and TLF exceeds 28\,fm.

To solve the quantal diffusion equations, we need to define the window plane which divides
the colliding system into two parts. We place the window plane at the minimum density location
along the elongation axis, as in Refs.~\cite{Yilmaz(2014)1,Ayik(2015)2,Ayik(2018),Ayik(2019)1}
(i.e., the window moves as a function of time). For a better detection of the minimum density,
we first define $\rho(x,y)=\int \rho(\bs{r})dz$, integrated over the axis perpendicular to
the reaction plane, and then use the fifth-order polynomial interpolations for $x$ and $y$
directions. See Fig.~\ref{Fig:rho(t)}, which depicts a typical example of the reaction dynamics,
and also shows the window plane (indicated by a red line in each frame). To evaluate the neutron
and proton numbers of the projectile-like subsystem, $N_1(t)$ and $Z_1(t)$, a smooth step-like
function, $\Theta(x') = \frac{1}{2}[1 + \tanh(\alpha x'/\Delta x)]$, is used, where $|x'|$
is the distance from the window plane and $\Delta x$ is the mesh spacing, with $\alpha=3$.
For the memory integral in the quantal diffusion coefficient, Eq.~(\ref{Eq:D_qq}), we
replace $\int_0^t d\tau$ with $\int_0^T d\tau$, setting $T$\,$=$\,0.33\,zs, which
is sufficiently long to include possible memory effects on the diffusion process.
We have confirmed that the memory effects in the first term of Eq.~(\ref{Eq:D_qq})
can be well approximated with neglecting the memory time dependence of the currents,
i.e., with $\int_0^tG_\mu(\tau)d\tau\approx1/2$.
The details of the determination of the curvature parameters for the driving potential in
Eqs.~(\ref{Eq:nu_n_Einstein}) and (\ref{Eq:nu_p_Einstein}) are given in Appendix~\ref{Appendix1}

For the particle-number projection method for TDHF [Eq.~(\ref{Eq:P_NZ_TDHF})], the interval
$[0,2\pi]$ is discretized into 300 uniform grids. Since the experiments \cite{Krolas(2003),Krolas(2010)}
were carried out with thick targets and reaction products were identified by subsequent
decay properties, the data should contain information of fragments in the whole angular
range. To include all contributions for transfer products, we include impact parameters of
$b$\,$\lesssim$\,10\,fm, which correspond to the orbital angular momentum range up
to $L$\,$=$\,240$\hbar$, where $L$\,$=$\,$b(2\mu E_{\rm c.m.})^{1/2}$ (in units of $\hbar$)
with the reduced mass $\mu$. We mention here that for the $^{64}$Ni+$^{208}$Pb
system with $L$\,$<$\,30$\hbar$ ($b$\,$\lesssim$\,1\,fm) we observed an abrupt change
of the minimum density location in the course of collision, which may be related to shell effects
mentioned in Sec.~\ref{sec:average}. We thus excluded $L$\,$<$\,30$\hbar$ from the
cross-section calculation, since it has little effect on the results. Also, for the $^{58}$Ni+$^{208}$Pb
system with $L$\,$=$\,45$\hbar$, we observed formation of binary products after a long
contact time, about 25\,zs. Since the contact time is rapidly increasing and the system is
on the border between fusion and binary reactions, $L$\,$=$\,45$\hbar$ is not included
in the cross-section calculation. \texttt{GEMINI++} calculations were carried out with the
default parameter set. We confirmed that the statistical treatment provides a convergent result.

\begin{figure} [t]
\includegraphics[height=9.cm]{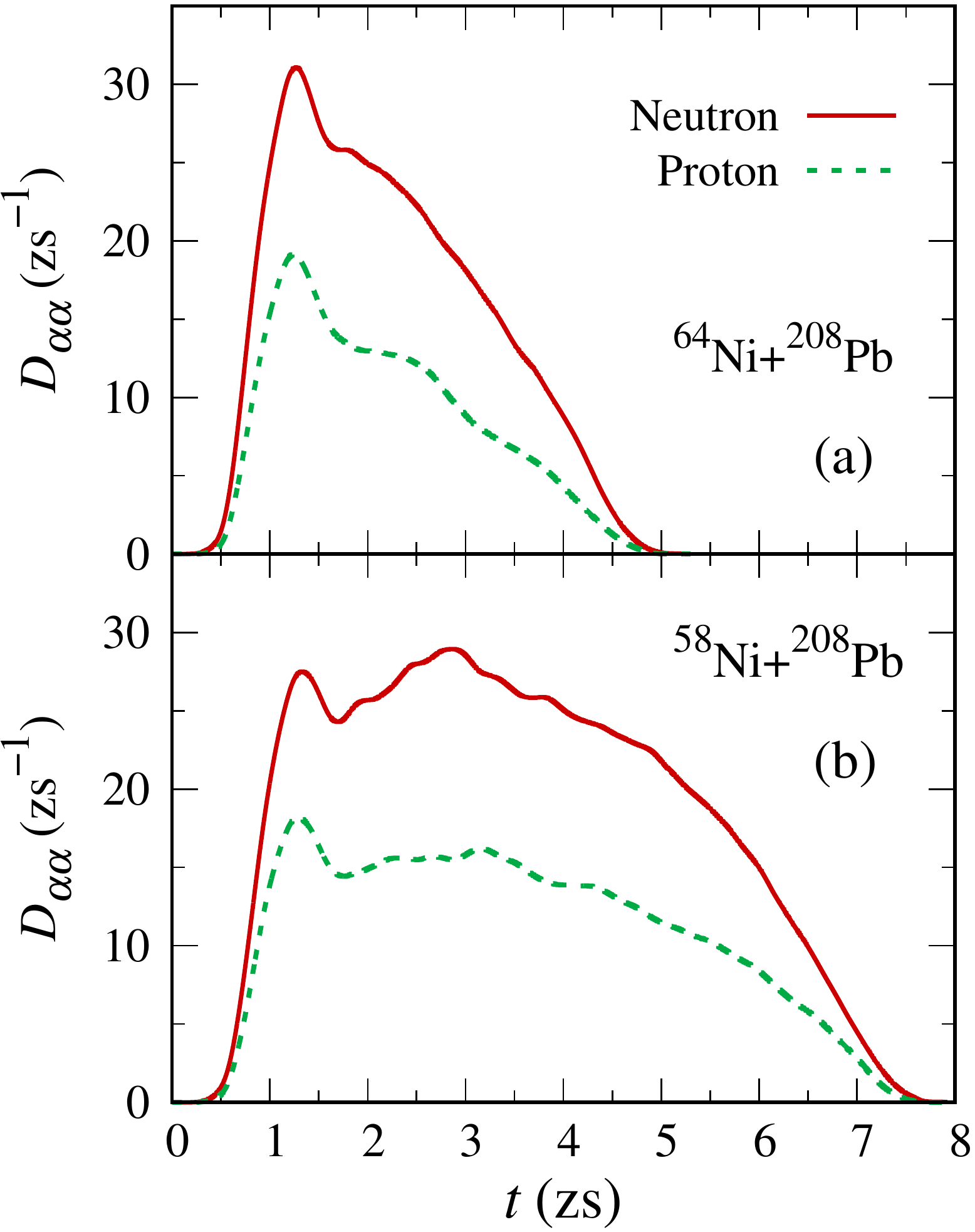}\vspace{-1mm}
\caption{
Diffusion coefficients for neutron and proton transfers are shown as a function of time.
Red solid line shows the diffusion coefficient for neutron transfer, $D_{NN}(t)$, while
green dashed line shows that for proton transfer, $D_{ZZ}(t)$. In panels (a) and (b),
results for the $^{64}$Ni+$^{208}$Pb reaction at $E_{\rm c.m.}$\,$=$\,268~MeV
and the $^{58}$Ni+$^{208}$Pb reaction at $E_{\rm c.m.}$\,$=$\,270~MeV are
presented, respectively. The initial orbital angular momentum is $L$\,$=$\,50$\hbar$
in both cases. Results were smoothed by taking an average over 0.67\,zs (see texts).
}\vspace{-3mm}
\label{Fig:D(t)}
\end{figure}

\begin{figure} [t]
\vspace{1.1mm}
\includegraphics[height=8.92cm]{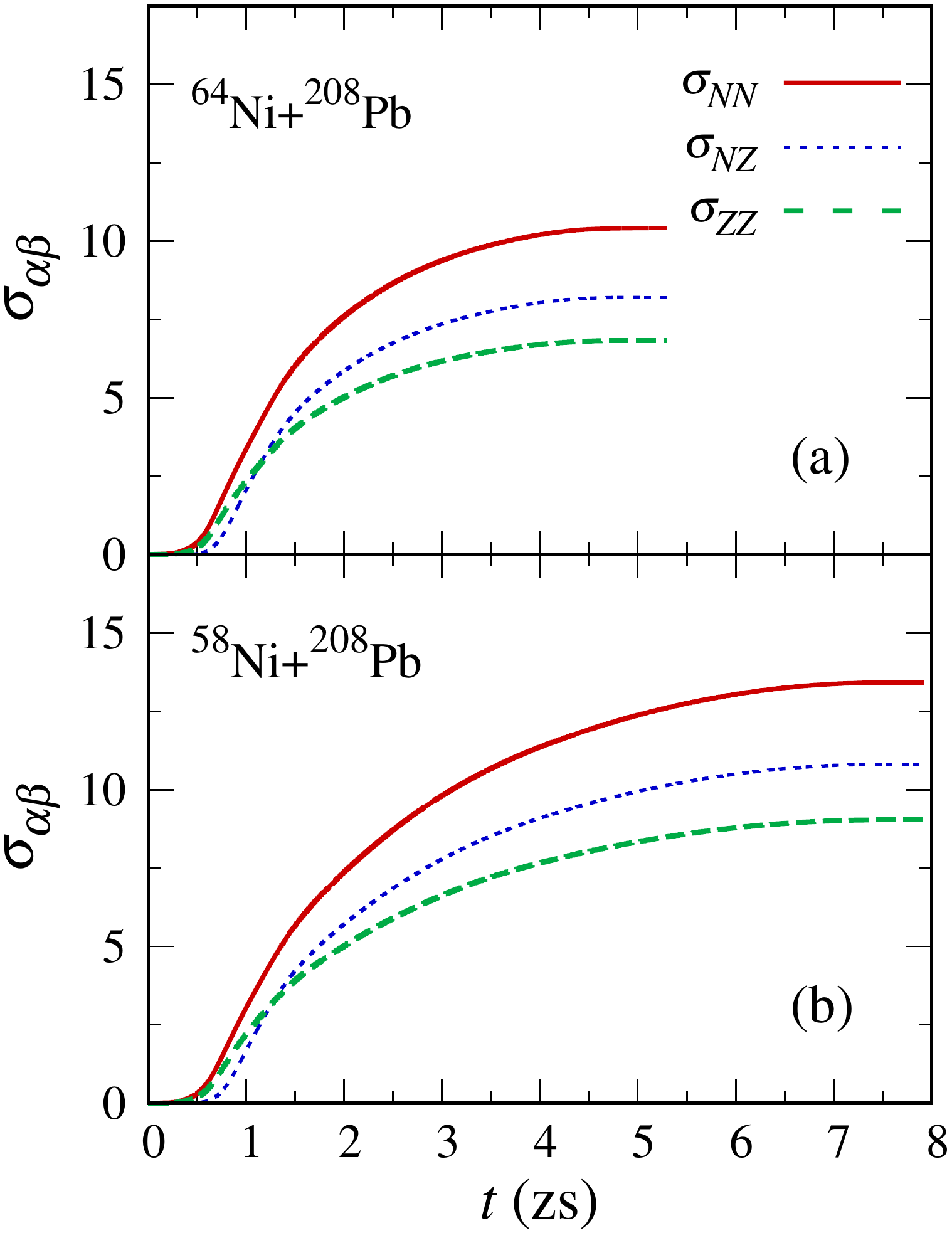}\vspace{-1mm}
\caption{
Fluctuations and correlation for nucleon transfers are shown as a function of time.
Red solid line shows the fluctuation in neutron transfer, $\sigma_{NN}(t)$, while green dashed
line shows that in proton transfer, $\sigma_{ZZ}(t)$. By blue dotted line, the correlation between
the neutron and proton transfers, $\sigma_{NZ}(t)$, is shown. In panels (a) and (b), results for the
$^{64}$Ni+$^{208}$Pb reaction at $E_{\rm c.m.}$\,$=$\,268~MeV and the $^{58}$Ni+$^{208}$Pb
reaction at $E_{\rm c.m.}$\,$=$\,270~MeV are presented, respectively. The initial orbital angular
momentum is $L$\,$=$\,50$\hbar$ in both cases.
}\vspace{-3mm}
\label{Fig:sigma_XY(t)}
\end{figure}

\begin{figure*} [t]
\begin{minipage}{\columnwidth}
\includegraphics[width=7.6cm]{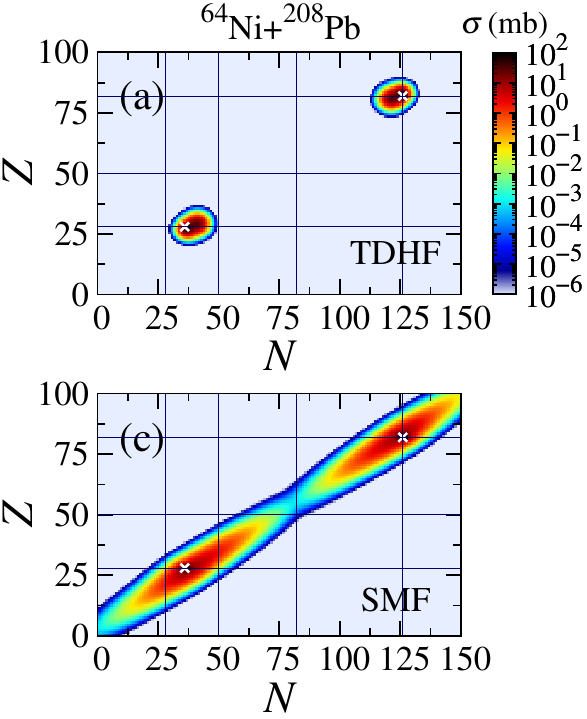}
\end{minipage}\;
\begin{minipage}{\columnwidth}
\includegraphics[width=7.6cm]{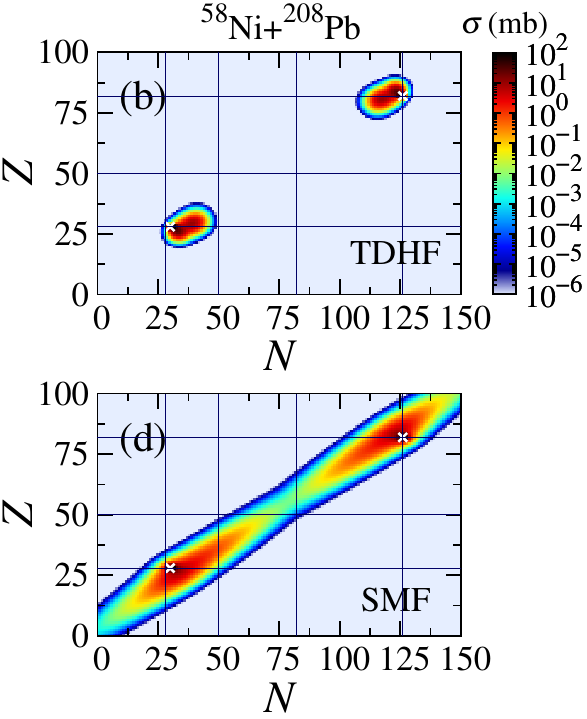}
\end{minipage}\vspace{-1mm}
\caption{
Primary production cross sections for both light (projectile-like) and heavy (target-like)
fragments are shown in the $N$-$Z$ plane for the $^{64}$Ni+$^{208}$Pb reaction
at $E_{\rm c.m.}$\,$=$\,268~MeV (left panels) and the $^{58}$Ni+$^{208}$Pb reaction
at $E_{\rm c.m.}$\,$=$\,270~MeV (right panels). Cross sections are shown in logarithmic scale
in the units of millibarn. Upper panels [Figs.~\ref{Fig:sigmatot(N,Z)}(a) and \ref{Fig:sigmatot(N,Z)}(b)]
show the results of TDHF calculations, while lower panels [Figs.~\ref{Fig:sigmatot(N,Z)}(c) and
\ref{Fig:sigmatot(N,Z)}(d)] show the results obtained by the quantal diffusion approach based
on the SMF theory. The horizontal lines indicate the proton magic numbers, $Z$\,$=$\,28, 50, and 82,
while the vertical lines indicate the neutron magic numbers, $N$\,$=$\,28, 50, 82, and 126. The crosses
point at the neutron and proton numbers of the projectile and target nuclei.
}\vspace{-3mm}
\label{Fig:sigmatot(N,Z)}
\end{figure*}

\section{RESULTS}\label{sec:results}

\subsection{Mean reaction dynamics in TDHF}\label{sec:average}

We analyze the multinucleon transfer mechanism in the $^{64}$Ni+$^{208}$Pb
reaction at $E_{\rm c.m.}$\,$=$\,268~MeV and the $^{58}$Ni+$^{208}$Pb
reaction at $E_{\rm c.m.}$\,$=$\,270~MeV. Those collision energies correspond to
$E_{\rm c.m.}/V_{\rm B}$\,$\simeq$\,1.13 and $E_{\rm c.m.}/V_{\rm B}$\,$\simeq$\,1.09
for the $^{64}$Ni+$^{208}$Pb and $^{58}$Ni+$^{208}$Pb systems, respectively,
where $V_{\rm B}$ denotes the Coulomb barrier height estimated with the frozen
Hartree-Fock method \cite{Simenel(2017)}. As a typical example of the reaction dynamics,
time evolution of the density in the $^{64}$Ni+$^{208}$Pb reaction at $E_{\rm c.m.}$\,$=$\,268~MeV
with $L$\,$=$\,50$\hbar$ is shown in Fig.~\ref{Fig:rho(t)}. Density contour plots at typical instances
are shown in the reaction plane, where the elongation axis and the window plane are indicated
as well. A great advantage of the TDHF approach is that it provides us intuitive information
on the time evolution of nuclear dynamics, which is not a direct observable in experiments
(See Ref.~\cite{SupplementalMovies} for full movies of the reactions within the TDHF approach).
In the course of the reaction, two nuclei collide deeply and then form a dinuclear structure
connected with a thick neck (see, $t$\,$=$\,2.67\,zs) that allows us to apply the quantal
diffusion description for multinucleon transfers (cf. Sec.~\ref{Sec:diffusion}). A number of
nucleons are exchanged between two subsystems through the window. After substantial diffusion
of nucleon numbers, the dinuclear system eventually splits into two (see, $t$\,$=$\,4.48\,zs).
The well separated binary reaction products (see, $t$\,$=$\,5.3\,zs) are then analyzed and
various observables are calculated. Numerical results are presented in Tables~\ref{Table:64Ni+208Pb}
and \ref{Table:58Ni+208Pb} in Appendix~\ref{Appendix:Tables}.

In Figs.~\ref{Fig:TKEL} and \ref{Fig:N_and_Z}, open circles show the results of TDHF calculations
for the $^{64}$Ni+$^{208}$Pb reaction, while open squares show those for the $^{58}$Ni+$^{208}$Pb
reaction. In Figs.~\ref{Fig:TKEL}(a) and \ref{Fig:TKEL}(b), we show, respectively, the contact
time and the average TKEL as a function of the initial orbital angular momentum. The contact
time is defined as a duration during which the minimal density between colliding nuclei exceeds
half the nuclear saturation density, $\rho_{\rm sat}/2$\,$\simeq$\,0.08\,fm$^{-3}$. From
the figure, we see a sharp increase of TKEL as the orbital angular momentum decreases, where
the contact time becomes finite. In both collisions the maximum amount of TKEL is about 65--70~MeV,
which occurs for the initial angular momentum less than $L$\,$\le$\,100$\hbar$ for the
$^{64}$Ni+$^{208}$Pb system and $L$\,$\le$\,90$\hbar$ for the $^{58}$Ni+$^{208}$Pb system.

In Fig.~\ref{Fig:TKEL}(c), we show the mean values of the charge asymmetry of the primary
PLF and TLF, $\delta=(N-Z)/(N+Z)$, as a function of the contact time. Note that each point
corresponds to the result with different orbital angular momenta. The equilibrium values
of the charge asymmetries of the composite dinuclear systems are $\delta$\,$\simeq$\,0.19
and 0.17, for the $^{64}$Ni+$^{208}$Pb and $^{58}$Ni+$^{208}$Pb systems, respectively.
These values are indicated by horizontal dotted lines in the figure. As can be seen from
Fig.~\ref{Fig:TKEL}(c), a fast charge equilibration process takes place within about 1~zs.
Then, the system evolves slowly toward the mass equilibrium, keeping the isospin asymmetry
roughly constant. Note that the systems do not reach the equilibrium values, because the
systems do not reach the mass equilibrium as well. The saturated values of $\delta$ correspond
to the $N/Z$ ratios of about 1.4 and 1.5 for PLF and TLF, respectively, in $^{64}$Ni+$^{208}$Pb
and 1.46 for TLF in $^{58}$Ni+$^{208}$Pb, which are in good agreement with the experimentally
deduced average $N/Z$ ratios \cite{Krolas(2003),Krolas(2010)}. 

\begin{figure*} [t]
\includegraphics[width=0.85\textwidth]{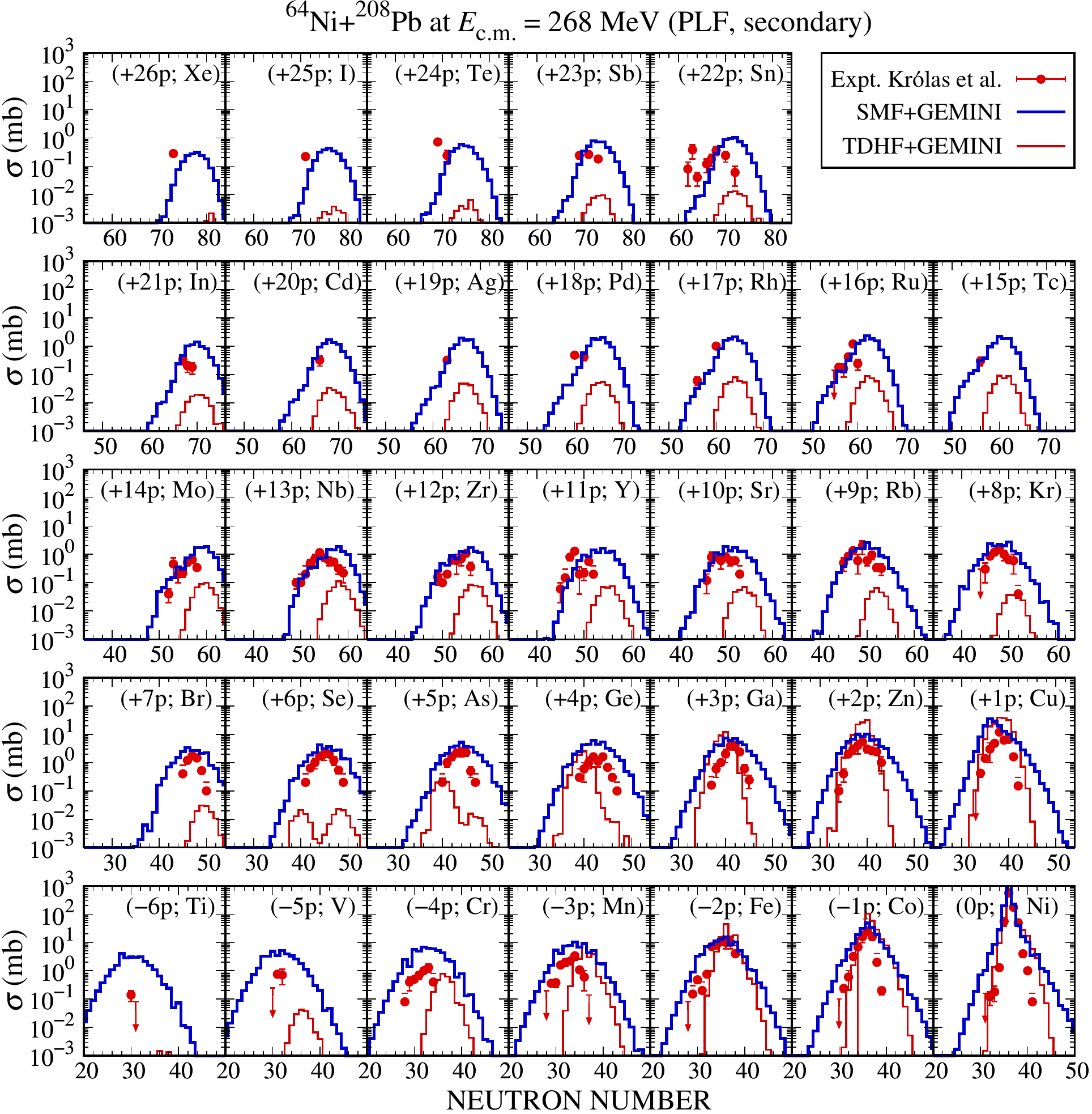}\vspace{-1mm}
\caption{
Secondary production cross sections for lighter (projectile-like) fragments in the $^{64}$Ni+$^{208}$Pb
reaction at $E_{\rm c.m.}$\,$=$\,268~MeV. In each panel, production cross sections for different isotopes
[as indicated by ($\pm x$p;~X), where $x$ indicates the number of transferred protons and X stands for
the corresponding element] are shown as a function of the neutron number. Blue thick histograms show
the results of SMF+GEMINI calculations, while red thin histograms show those of TDHF+GEMINI.
Red solid circles show the experimental data with error bars, taken from Ref.~\cite{Krolas(2003)}.
The down arrows indicate upper bound of the measured cross sections \cite{Krolas(2003)}.
}\vspace{-3mm}
\label{Fig:sigmatot_secondary_PLF_64Ni+208Pb}
\end{figure*}

\begin{figure*} [t]
\includegraphics[width=0.85\textwidth]{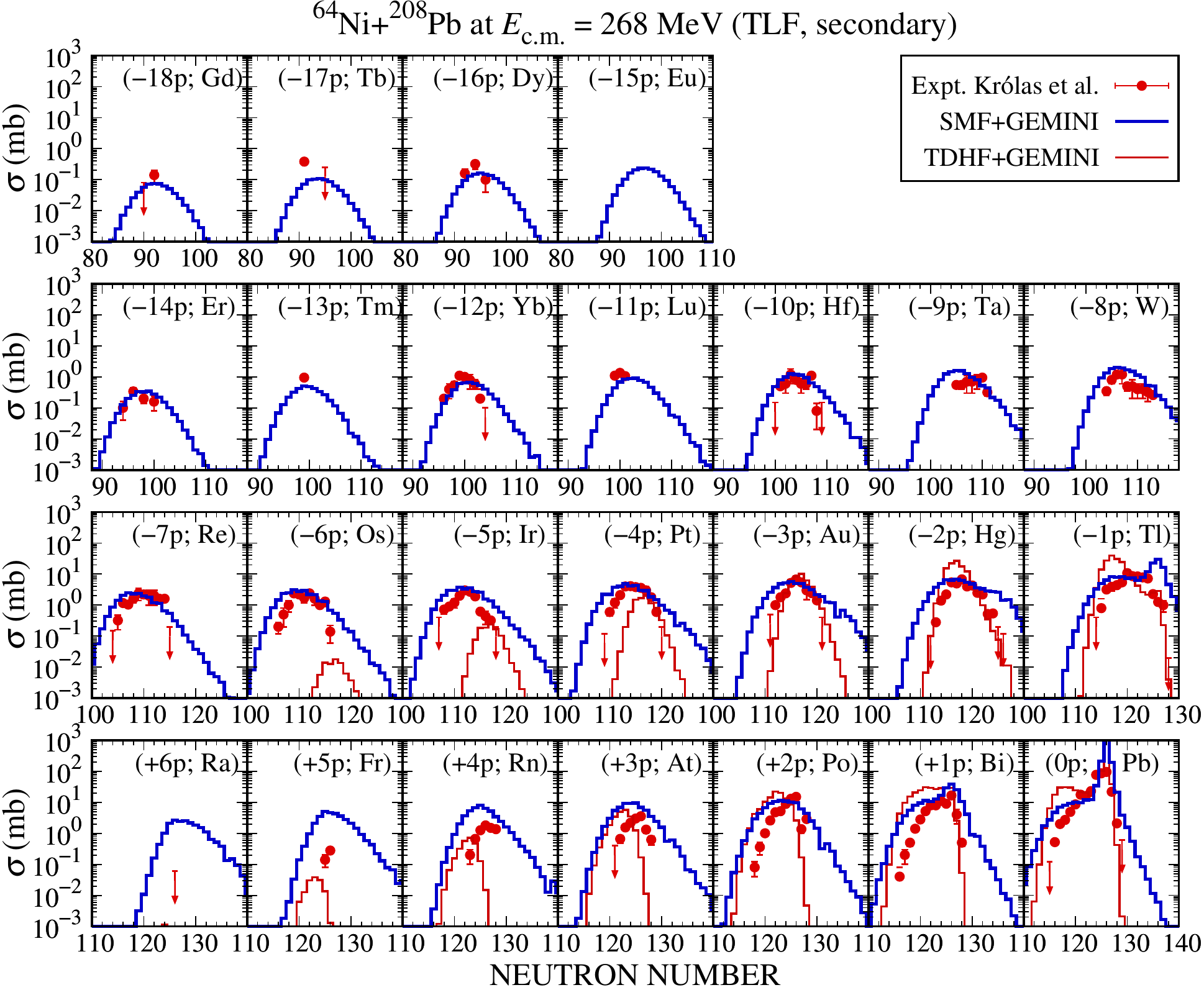}\vspace{-1mm}
\caption{
Same as Fig.~\ref{Fig:sigmatot_secondary_PLF_64Ni+208Pb}, but for heavier (target-like) fragments.
}\vspace{-3mm}
\label{Fig:sigmatot_secondary_TLF_64Ni+208Pb}
\end{figure*}

\begin{figure*} [t]
\includegraphics[width=0.85\textwidth]{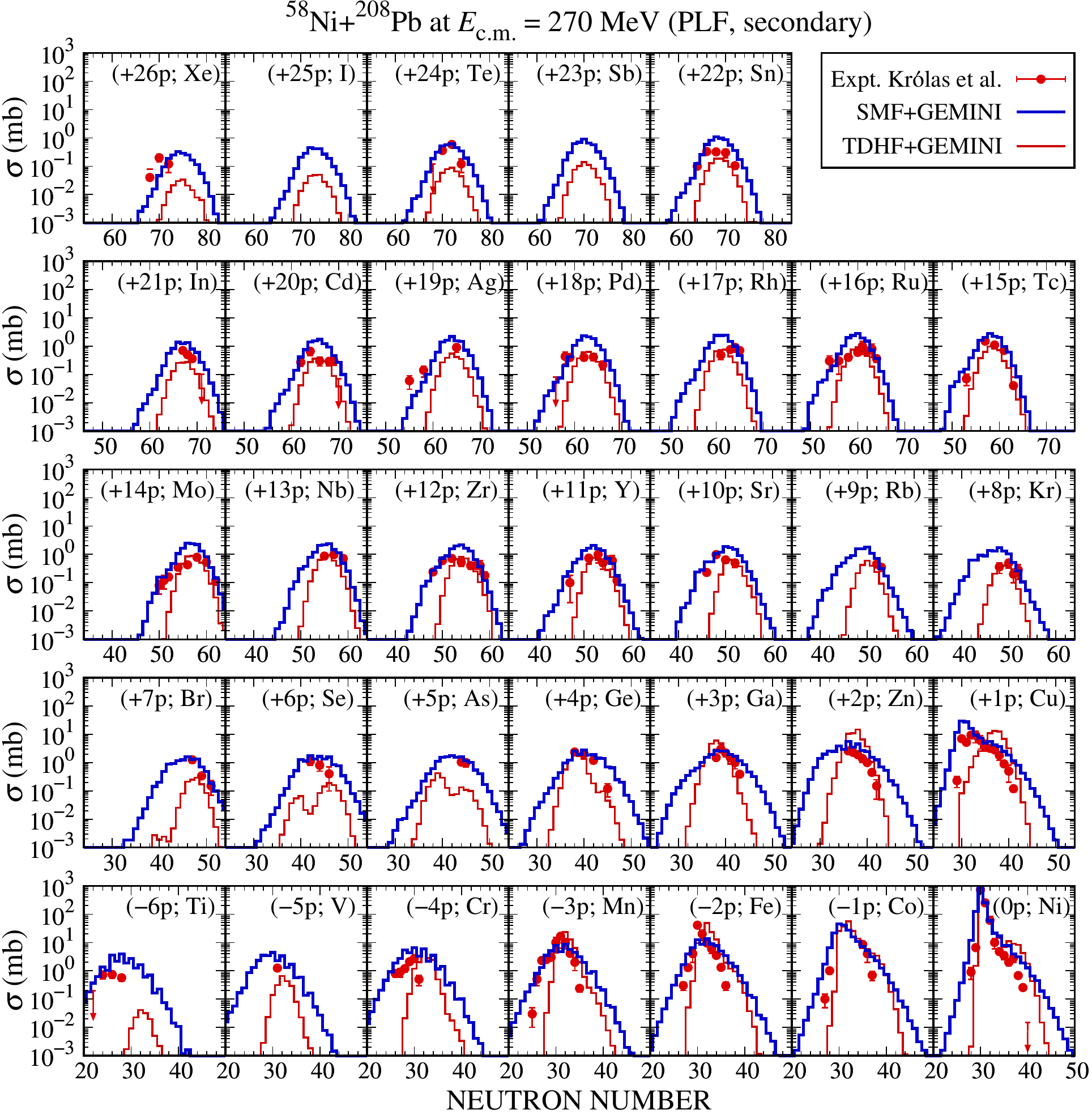}\vspace{-1mm}
\caption{
Same as Fig.~\ref{Fig:sigmatot_secondary_PLF_64Ni+208Pb}, but for the $^{58}$Ni+$^{208}$Pb
reaction at $E_{\rm c.m.}$\,$=$\,270~MeV. The experimental data were taken from Ref.~\cite{Krolas(2010)}.
}\vspace{-3mm}
\label{Fig:sigmatot_secondary_PLF_58Ni+208Pb}
\end{figure*}

\begin{figure*} [t]
\includegraphics[width=0.85\textwidth]{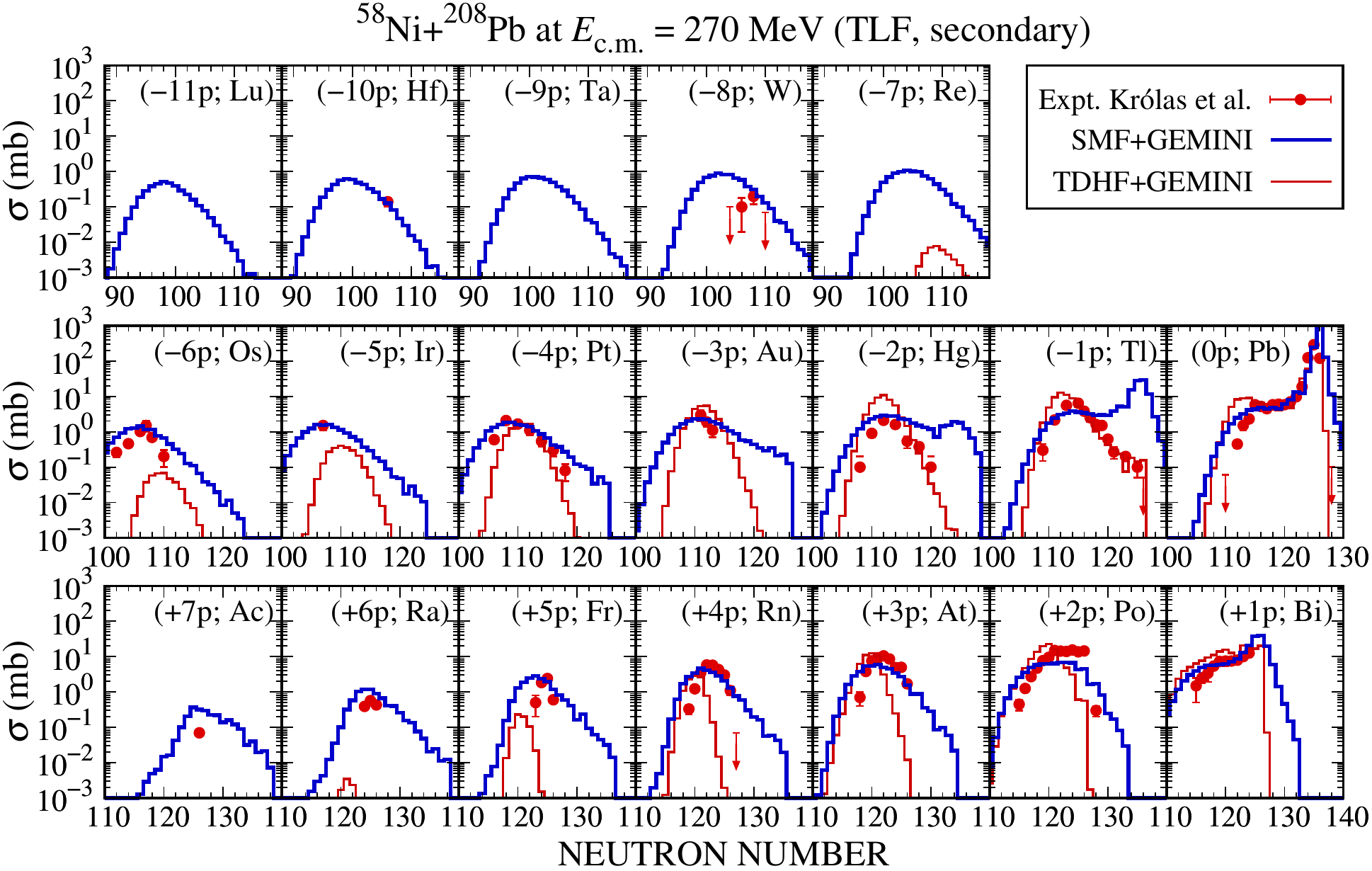}\vspace{-1mm}
\caption{
Same as Fig.~\ref{Fig:sigmatot_secondary_PLF_58Ni+208Pb}, but for heavier (target-like) fragments.
}\vspace{-3mm}
\label{Fig:sigmatot_secondary_TLF_58Ni+208Pb}
\end{figure*}

To provide more detailed information on nucleon transfers, in Fig.~(\ref{Fig:N_and_Z}),
we show the average number of protons (left panels) and neutrons (right panels) in reaction
products as a function of the initial orbital angular momentum. The upper panels
[Figs.~\ref{Fig:N_and_Z}(a) and \ref{Fig:N_and_Z}(b)] show those of heavier (target-like)
fragments, $Z_{\rm H}$ and $N_{\rm H}$, while the lower panels [Figs.~\ref{Fig:N_and_Z}(c)
and \ref{Fig:N_and_Z}(d)] show those of lighter (projectile-like) fragments, $Z_{\rm L}$
and $N_{\rm L}$. For the quasielastic regime with large orbital angular momenta
($L$\,$\gtrsim$\,150$\hbar$), where TKEL and contact time are almost zero, the average
neutron and proton numbers of the reaction products coincide with the initial values. As the
orbital angular momentum decreases, two nuclei touche in the course of the collision and a rapid
charge equilibration process takes place. At this stage, neutrons and protons are transferred
toward the opposite directions as expected from the initial isospin asymmetries, that is,
$^{208}$Pb\,$\rightarrow$\,$^{58,64}$Ni for neutrons, and $^{58,64}$Ni\,$\rightarrow$\,$^{208}$Pb
for protons. The latter process is relatively fast, which occurs within 1\,zs [cf. Fig.~\ref{Fig:TKEL}(a)],
governed by nucleons around the Fermi level. As the orbital angular momentum decreases further,
the dinuclear system starts evolving toward the mass symmetry. This trend is visible for
$80\hbar$\,$\lesssim$\,$L$\,$\lesssim$\,$120\hbar$ and $50\hbar$\,$\lesssim$\,$L$\,$\lesssim$\,$90\hbar$
for the $^{64}$Ni+$^{208}$Pb and $^{58}$Ni+$^{208}$Pb systems, respectively. We find,
however, that at small orbital angular momenta ($L$\,$<$\,80$\hbar$) the mass
equilibration process is terminated in the $^{64}$Ni+$^{208}$Pb reaction. It is likely
due to the shell effects around $Z_{\rm H}=82$ and $Z_{\rm L}=28$, which are weakened
by the fast isospin equilibration process in the $^{58}$Ni+$^{208}$Pb case. Notice that the
contact time increases more rapidly for $^{58}$Ni+$^{208}$Pb as compared to $^{64}$Ni+$^{208}$Pb
[see Fig.~\ref{Fig:TKEL}(a)], indicating that the former system favors to fuse. In fact, we observe
fusion reactions for $L$\,$\le$\,40$\hbar$ in the $^{58}$Ni+$^{208}$Pb reaction, where
the system does not splits for more than 40\,zs, whereas no fusion was observed in the
$^{64}$Ni+$^{208}$Pb reaction. (We mention here that in Ref.~\cite{KS_KY_MNT} the
$^{58}$Ni+$^{208}$Pb reaction at slightly lower energy, $E_{\rm c.m.}$\,$=$\,257~MeV,
was investigated in TDHF and very similar reaction dynamics were observed.)

\subsection{Quantal diffusion for multinucleon transfers}

In Fig.~\ref{Fig:D(t)}, we show the diffusion coefficients for neutron and proton transfers,
$D_{NN}(t)$ (solid line) and $D_{ZZ}(t)$ (dashed line), respectively, as a function of time. The
diffusion coefficients are evaluated according to Eq.~(\ref{Eq:D_qq}), which are entirely determined by
time evolution of the occupied single-particle orbitals in TDHF, as described in Sec.~\ref{Sec:TransportCoef}.
To eliminate rapid oscillations due to complex dynamics of single-particle degrees of freedom including
shell effects, the diffusion coefficients are smoothed by taking an average over 0.67\,zs, as in
the case of the mean drift path described in Appendix~\ref{Appendix1}. In Figs.~\ref{Fig:D(t)}(a)
and \ref{Fig:D(t)}(b), the results for the $^{64}$Ni+$^{208}$Pb and $^{58}$Ni+$^{208}$Pb
reactions are presented, respectively. The initial orbital angular momentum is $L$\,$=$\,50$\hbar$
for both cases (see Fig.~\ref{Fig:rho(t)} for the corresponding density plots in the $^{64}$Ni+$^{208}$Pb
reaction). From the figure, it is clear that the diffusion coefficient for neutron transfers is systematically
larger than that for protons, indicating influence of the Coulomb repulsion. In the initial stage of
the reaction up to $t$\,$\lesssim$\,1.5\,zs, both systems exhibit very similar behavior of the
diffusion coefficients. The maximum values of the diffusion coefficients are around 50\,zs$^{-1}$
for neutrons and 30\,zs\,$^{-1}$ for protons. As mentioned above, the contact time is longer
for the $^{58}$Ni+$^{208}$Pb system, in particular since $L$\,$=$\,50$\hbar$ is close to the
border between fusion and binary events. As a result, the diffusion coefficients extend for
a longer period for $^{58}$Ni+$^{208}$Pb [Fig.~\ref{Fig:D(t)}(b)] as compared to
$^{64}$Ni+$^{208}$Pb [Fig.~\ref{Fig:D(t)}(a)].

In Fig.~\ref{Fig:sigma_XY(t)}, the time evolution of the fluctuations and the correlation in neutron
and proton transfers are shown as functions of time. Again, the results for the $^{64}$Ni+$^{208}$Pb
and $^{58}$Ni+$^{208}$Pb reactions are shown in panels (a) and (b), respectively. The time
evolution is obtained by solving the set of partial differential equations, Eqs.~(\ref{Eq:sigma2_NN})--(\ref{Eq:sigma2_NZ}).
In the initial stage of the reaction up to $t$\,$\lesssim$\,1.25\,zs, we find that the magnitude
orders as $\sigma_{NZ}$\,<\,$\sigma_{ZZ}$\,<\,$\sigma_{NN}$. As time evolves further,
the correlation develops, changing the order as $\sigma_{ZZ}$\,<\,$\sigma_{NZ}$\,<\,$\sigma_{NN}$,
indicating the importance of correlations after substantial energy dissipation [cf. Figs.~\ref{Fig:TKEL}(a)
and \ref{Fig:TKEL}(b)]. We note that the correlation $\sigma_{NZ}$ is strictly zero within the TDHF
approach. The magnitude of the fluctuations in SMF is much larger than that obtained in TDHF. For
instance, in TDHF we find $\sigma_{NN}^{\rm TDHF}$\,$=$\,1.43 and $\sigma_{ZZ}^{\rm TDHF}$\,$=$\,1.27
for $^{64}$Ni+$^{208}$Pb, and $\sigma_{NN}^{\rm TDHF}$\,$=$\,1.58 and $\sigma_{ZZ}
^{\rm TDHF}$\,$=$\,1.38 for $^{58}$Ni+$^{208}$Pb, at $L$\,$=$\,50$\hbar$. Clearly, TDHF
severely underestimates the magnitude of the fluctuations and the correlation in dissipative collisions.

\subsection{Primary production cross sections}\label{sec:primary}

In Fig.~\ref{Fig:sigmatot(N,Z)}, we show the primary production cross sections $\sigma(N,Z)$
in the $N$-$Z$ plane for the $^{64}$Ni+$^{208}$Pb reaction at $E_{\rm c.m.}$\,$=$\,268~MeV
(left panels) and the $^{58}$Ni+$^{208}$Pb reaction at $E_{\rm c.m.}$\,$=$\,270~MeV (right panels).
In the upper panels [Figs.~\ref{Fig:sigmatot(N,Z)}(a) and \ref{Fig:sigmatot(N,Z)}(b)] the results
of TDHF calculations are shown, while the lower panels [Figs.~\ref{Fig:sigmatot(N,Z)}(c) and
\ref{Fig:sigmatot(N,Z)}(d)] show the results obtained within the quantal diffusion approach.
The figure clearly exhibits the fact that TDHF indeed provides quite narrow distributions in both systems.
In stark contrast, the SMF approach, as a result of the quantal diffusion mechanism, describes much
broader distributions and therefore it predicts production of a large number of primary fragments
in both systems. We note that the distributions in TDHF are nearly of round shape in the $N$-$Z$
plane, due to the product form of $P_{N,Z}(b)$ [Eq.~(\ref{Eq:P_NZ_TDHF})] without correlations
between neutron and proton transfers. The slightly skewed shape is due to the superposition of
contributions from different impact parameters. On the other hand, in the SMF approach, neutron
and proton transfers are substantially correlated, as indicated by a well-developed correlation
$\sigma_{NZ}$ in Fig.~\ref{Fig:sigma_XY(t)}. As a result, the distributions in SMF exhibit a strongly
correlated pattern which extends toward the mass symmetry of the system. The natural question is
if such a very wide distribution is realistic. We shall answer in the next section by comparing
the theoretical results with a full set of the experimental data reported by Kr\'olas \textit{et al.}
in Refs.~\cite{Krolas(2003),Krolas(2010)}.

\subsection{Secondary production cross sections}\label{sec:secondary}

In Figs.~\ref{Fig:sigmatot_secondary_PLF_64Ni+208Pb} and \ref{Fig:sigmatot_secondary_TLF_64Ni+208Pb},
we show the secondary production cross sections for PLFs and TLFs, respectively, in the $^{64}$Ni+$^{208}$Pb
reaction at $E_{\rm c.m.}$\,$=$\,268~MeV and compare the results with the measured cross sections \cite{Krolas(2003)}.
Similarly, in Figs.~\ref{Fig:sigmatot_secondary_PLF_58Ni+208Pb} and \ref{Fig:sigmatot_secondary_TLF_58Ni+208Pb},
we show the secondary production cross sections for PLFs and TLFs in the $^{58}$Ni+$^{208}$Pb reaction at
$E_{\rm c.m.}$\,$=$\,270~MeV, respectively, with the experimental data \cite{Krolas(2010)}. In those figures,
the measured cross sections are shown by solid circles with error bars. Upper bound of the measured cross sections
\cite{Krolas(2003),Krolas(2010)} is also indicated by a down arrow when available. The results obtained by
the quantal diffusion approach based on the SMF theory are shown by thick histograms, while the results
of TDHF calculations are shown by thin histograms. Each panel shows the isotopic distribution as a function
of the neutron number of the reaction product. The number of transferred protons and the corresponding
element are indicated by ($\pm x$p,~X), where the sign indicates difference relative to the proton number
of the projectile or the target and X stands for the corresponding element.

We start with a discussion on the TDHF results (red thin histograms, denoted by ``TDHF+GEMINI'' in the
figures), which are expected to work well around the average values and do not include the correlation
between neutron and proton transfers. First, we focus on the channels accompanying proton transfer from
nickel to lead, corresponding to the direction of the isospin equilibration in the initial systems, $^{58,64}$Ni
and $^{208}$Pb (the bottom row of the figures). The latter channels are mainly contributed from the
fast charge equilibration process at peripheral (grazing) collisions [cf. Figs.~\ref{Fig:TKEL} and \ref{Fig:N_and_Z}].
This type of processes is ubiquitous in systems with relatively large isospin asymmetry (see, e.g.,
Refs.~\cite{KS_KY_MNT,Corradi(40Ca+124Sn),Corradi(64Ni+238U),Szilner(40Ca+208Pb),Corradi(58Ni+208Pb),
Mijatovic(2016)}). From the figures, one can see that TDHF works fine when the number of transferred nucleons
is small [e.g., (0p) channel with transfer of several neutrons, ($-1$p) and ($-2$p) channels with transfer
of a few neutrons in Fig.~\ref{Fig:sigmatot_secondary_PLF_64Ni+208Pb}]. However, as the number of
transferred protons increases further [see, e.g., ($-$3p)--($-$6p) channels in Fig.~\ref{Fig:sigmatot_secondary_PLF_64Ni+208Pb}],
TDHF substantially underestimates the magnitude of the production cross sections. Moreover, the peak
position of the isotopic distributions for PLFs appears too neutron-rich [see, e.g., ($-$3p)--($-$5p) channels
in Fig.~\ref{Fig:sigmatot_secondary_PLF_64Ni+208Pb}]. This trend was also observed for other systems
\cite{KS_GEMINI}, although it was not clear if this is due to underestimation of neutron evaporation
effects or not. Interestingly, by looking at the corresponding proton-transfer channels for the heavier
parter [i.e., ($+$3p)--($+$5p) in Fig.~\ref{Fig:sigmatot_secondary_TLF_64Ni+208Pb}], we find that
the peak position of the isotopic distributions for TLFs appears too neutron-deficient. The combination
of those two observations indicates that proton removal (addition) tends to accompany neutron removal
(addition), implying the importance of the correlation in neutron and proton transfers.

Next, we discuss the TDHF results for channels accompanying transfer toward the direction of the mass
symmetry. As discussed in Sec.~\ref{sec:average}, for small initial orbital angular momenta, an onset of
quasifission emerges, where the system starts evolving toward the mass symmetry. The latter process
corresponds to the proton pickup ($+x$p) with respect to the projectile (Figs.~\ref{Fig:sigmatot_secondary_PLF_64Ni+208Pb}
and \ref{Fig:sigmatot_secondary_PLF_58Ni+208Pb}) and to the proton removal ($-x$p) with respect to
the target (Figs.~\ref{Fig:sigmatot_secondary_TLF_64Ni+208Pb} and \ref{Fig:sigmatot_secondary_TLF_58Ni+208Pb}).
Because of relatively short contact times, however, the system does not reach the mass equilibrium and we
observed transfer of only two protons from Ni to Pb on average in TDHF [cf. Figs.~\ref{Fig:N_and_Z}(a)
and \ref{Fig:N_and_Z}(c)]. As a result, TDHF provides a good description for channels with transfer of
a few protons from Ni to Pb, which are close to the average value [see, e.g., ($+$1p), ($+$2p), and ($+$3p)
channels in Fig.~\ref{Fig:sigmatot_secondary_PLF_64Ni+208Pb} and ($-$1p), ($-$2p), and ($-$3p)
channels in Fig.~\ref{Fig:sigmatot_secondary_TLF_64Ni+208Pb}]. As the number of transferred protons
increases further, however, TDHF substantially underestimates the magnitude of the production cross sections
[see, e.g., ($+$4p)--($+$7p) channels in Fig.~\ref{Fig:sigmatot_secondary_PLF_64Ni+208Pb} and
 and ($-$4p)--($-$7p) channels in Fig.~\ref{Fig:sigmatot_secondary_TLF_64Ni+208Pb}].
Furthermore, the experimental data for PLFs (Figs.~\ref{Fig:sigmatot_secondary_PLF_64Ni+208Pb}
and \ref{Fig:sigmatot_secondary_PLF_58Ni+208Pb}) exhibit considerable cross sections for channels
with transfer of a number of protons, up to ($+$26p) channel. We find that the cross sections for
those channels contain substantial contributions from transfer-induced fission of heavy partners.
Comparing with the experimental data, the magnitude of those cross sections are underestimated for
$^{64}$Ni+$^{208}$Pb (Fig.~\ref{Fig:sigmatot_secondary_PLF_64Ni+208Pb}), while it is
comparable to the data for $^{58}$Ni+$^{208}$Pb (Fig.~\ref{Fig:sigmatot_secondary_PLF_58Ni+208Pb}).
It indicates that the transfer-induced fission is reasonably simulated by \texttt{GEMINI++}.
On the other hand, for the TLFs, TDHF completely fails to reproduce the experimental data for
the channels accompanying removal of many protons [see, e.g., ($-$8p)--($-$18p) channels
in Fig.~\ref{Fig:sigmatot_secondary_TLF_64Ni+208Pb}]. This comparison clearly illustrates
the usefulness and limitation of TDHF+GEMINI in describing multinucleon transfer processes
in deep-inelastic collisions.

We shall now turn to a discussion on the results obtained by the quantal diffusion approach based on
the SMF theory (blue thick histograms, denoted by ``SMF+GEMINI'' in the figures). As can be seen from
Figs.~\ref{Fig:sigmatot_secondary_PLF_64Ni+208Pb}--\ref{Fig:sigmatot_secondary_TLF_58Ni+208Pb},
we find that the SMF approach provides a very good overall description of the measured production cross
sections, all the way up to ($+$26p) channel for lighter fragments (Figs.~\ref{Fig:sigmatot_secondary_PLF_64Ni+208Pb}
and \ref{Fig:sigmatot_secondary_PLF_58Ni+208Pb}) and ($-$18p) channel for heavier fragments
(Fig.~\ref{Fig:sigmatot_secondary_TLF_64Ni+208Pb}). Through an analysis of the data for PLFs
[Figs.~\ref{Fig:sigmatot_secondary_PLF_64Ni+208Pb} and \ref{Fig:sigmatot_secondary_PLF_58Ni+208Pb}],
we find that the production cross sections up to around ($+$14p) are dominated by the quantal
diffusion mechanism. In contrast, the production cross sections for Tc ($Z$\,$=$\,43) to Xe ($Z$\,$=$\,54)
isotopes [i.e., ($+$15p)--($+$26p) channels in Figs.~\ref{Fig:sigmatot_secondary_PLF_64Ni+208Pb}
and \ref{Fig:sigmatot_secondary_PLF_58Ni+208Pb}], we find a substantial contribution of transfer-induced
fission of heavy partners. It is remarkable that the SMF approach reproduces transfers in both directions,
owing to the quantal diffusion mechanism. The observed agreement between the SMF results and the
measurements suggests that a proper description for the production of heavy nuclei and their subsequent
decays is essential to reproduce the experimental data for the production of Tc to Xe isotopes.

We should point out, however, that the SMF approach overestimates the width of isotopic distributions,
see, e.g., the bottom two rows of Figs.~\ref{Fig:sigmatot_secondary_PLF_64Ni+208Pb}--\ref{Fig:sigmatot_secondary_TLF_58Ni+208Pb}. We consider that this overestimation is associated with the linearization of the Langevin equation (\ref{Eq:Langevin}).
In addition, there is disagreement with the experimental data, which is visible, for instance, in ($-$1p) channel
for TLF in the $^{64}$Ni+$^{208}$Pb system shown in Fig.~\ref{Fig:sigmatot_secondary_TLF_64Ni+208Pb}
and ($-$1p) and ($-$2p) channels for TLF in the $^{58}$Ni+$^{208}$Pb system shown in
Fig.~\ref{Fig:sigmatot_secondary_TLF_58Ni+208Pb}. Those channels are contributed by collisions
with large orbital angular momenta. We consider that the probability distribution of the Gaussian form
(\ref{Eq:P_NZ_SMF}) is inappropriate for this regime, as indicated by transfer probabilities in TDHF
\cite{Simenel(2010),KS_KY_MNT}. Indeed, for those channels TDHF correctly describes the shape
of the isotopic distributions. It is actually possible to further improve the quantal diffusion description
by incorporating nonlinear effects in the Langevin equation (\ref{Eq:Langevin}). Although we expect
that the improved quantal diffusion description provides a better description of the experimental data
presented in this work, we leave it as a future work.

\begin{figure} [t]
\includegraphics[width=8.0cm]{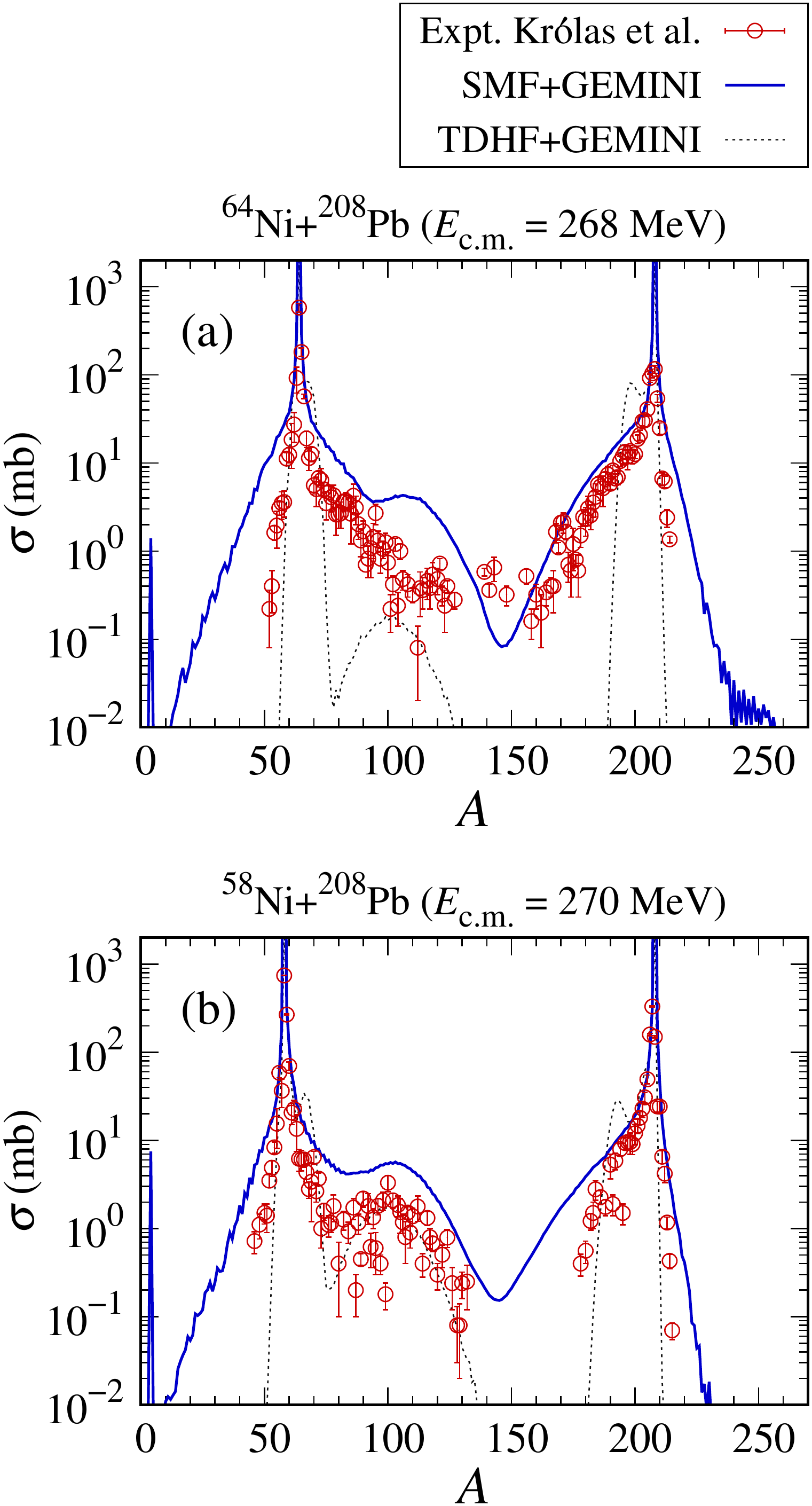}\vspace{-1mm}
\caption{
Mass distributions for secondary reaction products. In panels (a) and (b), results for
the $^{64}$Ni+$^{208}$Pb reaction at $E_{\rm c.m.}$\,$=$\,268~MeV and the
$^{58}$Ni+$^{208}$Pb reaction at $E_{\rm c.m.}$\,$=$\,270~MeV are shown,
respectively. The measured cross sections are shown by red open circles with error bars,
which were taken from Refs.~\cite{Krolas(2003),Krolas(2010)}. Blue solid line
represents the results of SMF+GEMINI calculations, while the results of TDHF+GEMINI
calculations are indicated by a thin dotted line.
}\vspace{-3mm}
\label{Fig:sigmatot(A)}
\end{figure}

For a complete representation of the data, we show in Fig.~\ref{Fig:sigmatot(A)} mass distributions
of secondary reaction products. In Fig.~\ref{Fig:sigmatot(A)}(a) and \ref{Fig:sigmatot(A)}(b),
the results for the $^{64}$Ni+$^{208}$Pb and $^{58}$Ni+$^{208}$Pb reactions are presented,
respectively. The experimental data \cite{Krolas(2003),Krolas(2010)} are shown by open circles
with error bars, while solid lines represents the results of SMF+GEMINI calculations. The results
of TDHF+GEMINI are also shown by thin dotted lines for comparison. We note that the data are
merely an integration of the absolute cross sections shown in
Figs.~\ref{Fig:sigmatot_secondary_PLF_64Ni+208Pb}--\ref{Fig:sigmatot_secondary_TLF_58Ni+208Pb}
for a given mass number $A$ and do not involve any renormalization for comparison. We also note
that the experimental data for the $^{58}$Ni+$^{208}$Pb reaction are less complete than the other,
since the off-line radioactivity measurement was not carried out for this system \cite{Krolas(2010)}.
From the figure we find that the SMF approach provides much wider mass distributions as compared
to the TDHF approach. A comparison with the SMF results and the measurements reveals that the
SMF approach predicts somewhat wider distributions, which are larger in magnitude, as compared
to the experimental data. The overestimation may partly be due to the linearization of the
Langevin equation, as mentioned above. We note, however, that the experimental data points in
Figs.~\ref{Fig:sigmatot_secondary_PLF_64Ni+208Pb}--\ref{Fig:sigmatot_secondary_TLF_58Ni+208Pb}
are limited in some regions, whereas the mass distributions in the SMF approach involve the
entire cross sections for all isotopes, that would result in larger magnitude. We note that there
is a sizable contribution of transfer-induced fission around $A$\,$\simeq$\,90--140 in both systems.
We find a noticeable contribution of transfer-induced fission also in TDHF+GEMINI, while it is
much bigger in SMF+GEMINI. Those fission products contribute to generate cross sections for
Tc to Xe isotopes [i.e., ($+$15p)--($+$26p) channels in Figs.~\ref{Fig:sigmatot_secondary_PLF_64Ni+208Pb}
and \ref{Fig:sigmatot_secondary_PLF_58Ni+208Pb}], as mentioned above.

\begin{figure} [t]
\includegraphics[width=8.0cm]{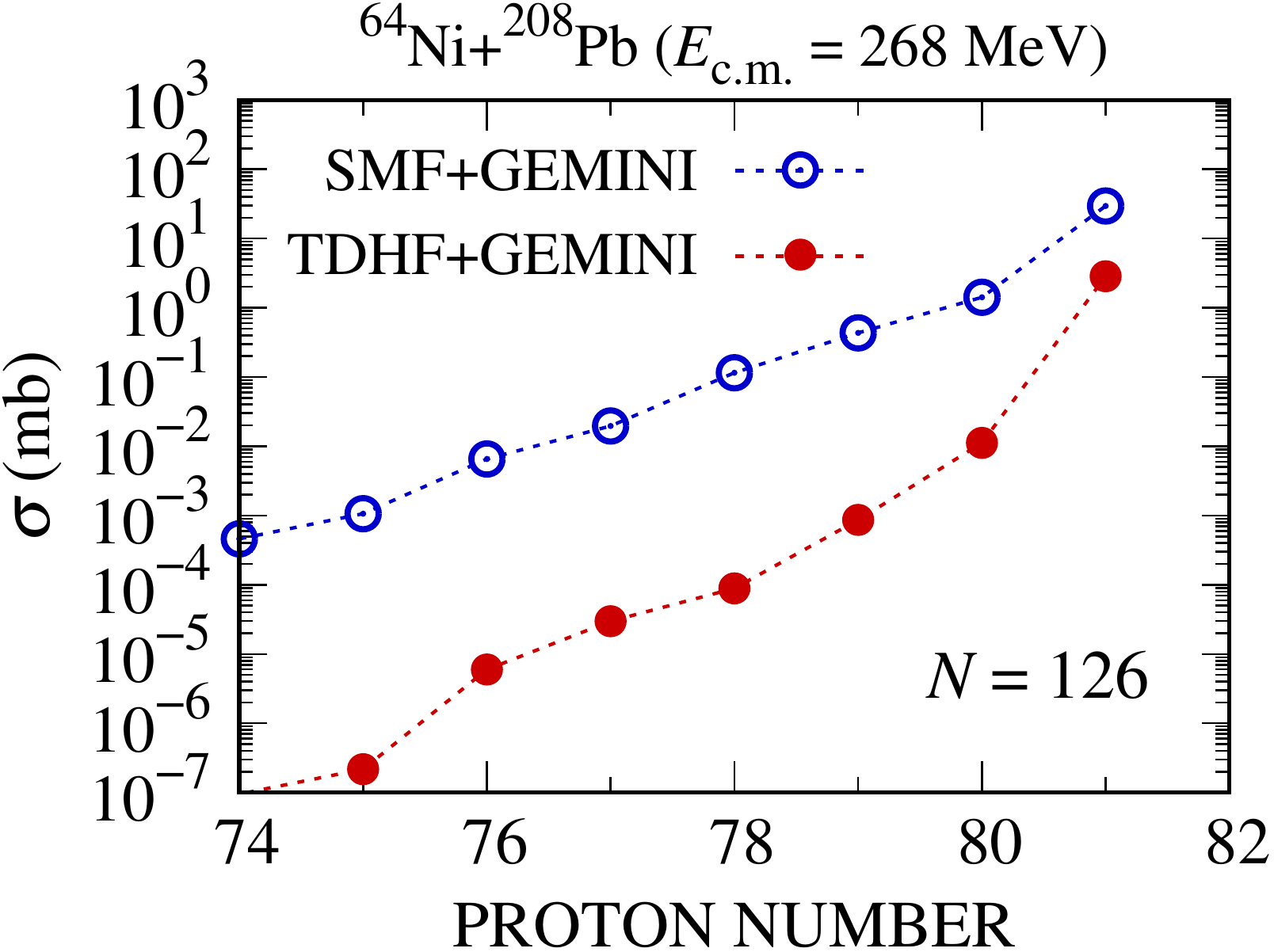}\vspace{-1mm}
\caption{
Secondary production cross sections for $N$\,$=$\,126 isotones in the $^{64}$Ni+$^{208}$Pb
reaction at $E_{\rm c.m.}$\,$=$\,268~MeV are shown as a function of the proton number
of the reaction products. Blue open circles represent the results of SMF+GEMINI calculations,
while red solid circles represent those of TDHF+GEMINI.
}\vspace{-3mm}
\label{Fig:sigmatot_N126}
\end{figure}

Finally, Fig.~\ref{Fig:sigmatot_N126} illustrates the prediction of the secondary production
cross sections for $N$\,$=$\,126 isotones in the $^{64}$Ni+$^{208}$Pb reaction at
$E_{\rm c.m.}$\,$=$\,268~MeV. In the figure, open and solid circles indicate the results
of the SMF and TDHF calculations, respectively. The magnitude of the cross sections are
closer for one-proton transfer, but the SMF approach predicts cross sections larger by
several orders of magnitude with increased number of removed protons from lead. We mention
here that the experimentally deduced cross sections for the $N$\,$=$\,126 isotones production
in the $^{136}$Xe+$^{198}$Pt reaction \cite{Watanabe(2015)} exhibit the magnitude
which is comparable to (or even slightly larger than) that indicated by SMF+GEMINI for
the present system. The figure highlights the necessity to go beyond the standard TDHF
description for a quantitative prediction of production of unstable nuclei with neutron
and proton numbers which are far apart from the average values.

\section{SUMMARY AND CONCLUSIONS}\label{sec:conclusions}

Recently, it has been hoped that the multinucleon transfer reaction in low-energy heavy-ion
reactions may be an efficient mechanism for production of yet-unknown neutron-rich
heavy nuclei and also for synthesizing neutron-rich superheavy elements. Aiming at
the production of new neutron-rich nuclei, much experimental effort has been undertaken
already that will be continued for the coming future. Several macroscopic or microscopic
transport approaches have been developed for theoretical investigations of the multinucleon
transfer mechanism \cite{Adamian(2020)}. These transport approaches are, in general,
very useful for analyzing the experimental data. However, because of a number of adjustable
parameters they involve, they have limited predictive power. The time-dependent Hartree-Fock
(TDHF) approach provides a microscopic description for low-energy heavy-ion reactions. While
the mean-field approach has been very successful for describing the most probable path
of the collective motion, it severely underestimates the dynamical fluctuations and
distributions of observables around their average values. Recently, as an extension of the
TDHF approach, the time-dependent random phase approximation (TDRPA) has been
applied to low-energy heavy-ion collisions. Since the TDRPA formula is obtained by linearizing
the equation of motion around the mean evolution, it provides a good approximation for
dispersion of one-body observables when amplitude of fluctuations is sufficiently small.
However, this approach appears to have a technical problem in describing dispersions of
one-body observables in asymmetric systems \cite{Williams(2018)}. As seen from recent
publications \cite{Ayik(2018),Yilmaz(2018),Ayik(2019)1,Ayik(2019)2,Yilmaz(2020)},
the quantal diffusion description for multinucleon exchanges based on the stochastic
mean-field (SMF) approach provides very good description for multinucleon transfer
mechanism, which is applicable for asymmetric reactions as well.

In this work, we have applied the quantal diffusion approach based on the SMF theory
to analyze the multinucleon transfer mechanism in the $^{64}$Ni+$^{208}$Pb reaction
at $E_{\rm c.m.}$\,$=$\,268~MeV and the $^{58}$Ni+$^{208}$Pb reaction at
$E_{\rm c.m.}$\,$=$\,270~MeV. The rich experimental data by Kr\'olas \textit{et al.}
\cite{Krolas(2003),Krolas(2010)}, which were measured with thick targets and
therefore include both projectile-like and target-like fragments as well as transfer-induced
fission products, allowed us to confirm the usefulness and limitations in the TDHF and the
SMF approaches. In general, it turns out that the quantal diffusion approach provides
very reasonable description of the experimental data, in both directions of transfers from
projectile to target and vice versa, for both projectile-like and target-like fragments.
A striking finding is a significant contribution of transfer-induced fission of heavy reaction
products, which nicely agree with the experimental observation. In some cases the SMF
calculations overestimate the isotopic widths, which may be due to the linear approximation
employed in the Langevin equation for macroscopic variables. Based on the quantal
diffusion approach, we have found that the production cross sections for $N$\,$=$\,126
isotones could be significantly larger than those expected within the TDHF approach
\cite{Jiang(2018),Jiang(2020),Jiang(2020)2}. It underlines the importance of going
beyond, especially to include one-body (mean-field) fluctuations and correlations,
the standard TDHF description for predicting production cross sections of unknown
neutron-rich heavy nuclei.

With the SMF theory we have achieved a remarkable progress in microscopic description
of low-energy heavy-ion reactions, as compared to the standard TDHF approach. Nevertheless,
there still remains some room for further improvements of the theoretical framework.
Below, we envision possible future directions:
\begin{enumerate}
\item\textit{Nonlinear effects in the quantal diffusion mechanism.}
As outlined in Sec.~\ref{Sec:diffusion}, in deriving the quantal diffusion description,
we have linearized the Langevin equation (\ref{Eq:Langevin}) around the mean trajectory,
assuming small amplitude fluctuations. Note that a similar assumption is made in the TDRPA
approach as well \cite{BV(1981),Simenel(review),Broomfield(2009),Simenel(2011),
Williams(2018),Godbey(2019)}. In addition, a simple parabolic from has been assumed
for the driving potential $U(N,Z)$ [cf. Eq.~(\ref{Eq:PES})], and it can be generalized by
introducing anharmonicity in the potential form. Such extensions are expected to reduce
the isotopic width and would result in better agreement with the experimental data.

\item\textit{TKE distributions.}
So far, we have not discussed how to describe distributions of the collective relative motion,
namely, fluctuations in scattering angles as well as total kinetic energy (TKE). The distribution
of TKE is, in turn, related to the total excitation energy distribution as discussed in Sec.~\ref{sec:TXE}.
In principle, one can adapt the SMF concept to the relative motion of a colliding system,
which allows us to evaluate fluctuations of, e.g., TKE based on microscopic mean-field
dynamics. A work is in progress along this line.

\item\textit{Pairing correlations.}
In the present work, we have neglected the pairing correlations, assuming that pairing in
atomic nuclei is so fragile and plays a minor role in dissipative heavy-ion collisions. However,
pairing may alter reaction dynamics in an unexpected way, as shown in, e.g., Refs.~\cite{
Hashimoto(2016),MSW(2017),SMW(2017),SWM(2017),Barton(2019)}. Actually, the SMF
approach has already been generalized to include the pairing correlations \cite{Lacroix(2013)}.
With the recent developments of mean-field approaches including pairing for static properties
\cite{Pei(2014),COCG(2017),Shi(2018),COCR(2020)} and dynamics \cite{Hashimoto(2016),
MSW(2017),SMW(2017),SWM(2017),Barton(2019),Ebata(2010),Scamps(2012),Scamps(2013),
Ebata(2015),Scamps(2017)}, the application of the quantal diffusion approach is feasible,
as outlined in this work.
\end{enumerate}

Finally, we comment on the Gaussian assumption of the initial fluctuations in the density
matrix, $\delta\rho_{ij}^\lambda$. In Refs.~\cite{Ulgen(2019),Yilmaz(2014)2}, effects
of relaxing of the Gaussian assumption have been explored. In particular, it has been shown
that third- and fourth-order moments of a one-body observable can be better described by an
appropriate relaxation \cite{Ulgen(2019)}. We note, however, that the Gaussian assumption
is used to formulate the quantal diffusion approach for multinucleon exchanges as described
in Sec.~\ref{Sec:diffusion}. Thus, although it is possible to further improve the model ingredients
for more realistic description including higher moments of one-body observables by relaxing
the Gaussian assumption, it is then necessary to generate an ensemble of TDHF trajectories
that requires vast computational costs. We also mention here that another possible extension
has been proposed recently in Ref.~\cite{Czuba(2019)}, where the SMF concept is applied
to the time-dependent reduced density matrix approach.

To conclude, in this work we have demonstrated that the quantal diffusion approach based
on the SMF theory is a powerful and promising tool of choice to microscopically and quantitatively
describe multinucleon transfer processes in low-energy heavy-ion reactions. We emphasize
that the quantal diffusion approach based on the SMF theory does not involve any adjustable
parameters, once an energy density functional is given, and transport coefficients are entirely
determined from the occupied single-particle orbitals in the TDHF approach. The observed
agreement with the full set of the experimental data is not only remarkable, but also
encouraging for microscopic mean-field theories for predicting and understanding various
outcomes from complex many-body dynamics of low-energy heavy-ion reactions.

\begin{acknowledgements}
The authors thank B\"ulent Yilmaz of Ankara University for useful comments on
the window dynamics. K.S. gratefully acknowledges Tennessee Technological
University for partial financial support and hospitality during his visit. S.A. is very
much thankful to his wife F.~Ayik for continuous support and encouragement.
This work is supported in part by U.S. Department of Energy (DOE) Grant No.\ DE-SC0015513,
and in part by JSPS Grant-in-Aid for Early-Career Scientists Grant No.\ 19K14704.
This work used the computational resource of the HPCI system (Oakforest-PACS)
provided by Joint Center for Advanced High Performance Computing (JCAHPC)
through the HPCI System Project (Project ID: hp190002), and (in part) by 
Multidisciplinary Cooperative Research Program (MCRP) in Center for Computational
Sciences (CCS), University of Tsukuba (Project ID: NUCLHIC). This work also used
(in part) computational resources of the Cray XC40 System at Yukawa Institute
for Theoretical Physics (YITP), Kyoto Univesity.
\end{acknowledgements}

\appendix
\setcounter{figure}{0}
\renewcommand{\thefigure}{\thesection\arabic{figure}}

\section{Derivatives of the mean drift coefficients}\label{Appendix1}

To solve the partial differential equations, Eqs.~(\ref{Eq:sigma2_NN})--(\ref{Eq:sigma2_NZ}),
one has to evaluate derivatives of the mean drift coefficients, $\nu_n$ and $\nu_p$, with respect
to the numbers of neutrons and protons in the projectile-like subsystem, $N_1$ and $Z_1$. In the
present article, we use the same strategy as in Refs.~\cite{Merchant(1981),Merchant(1982),Ayik(2015)2}.
For completeness, we provide below the details of the procedure for the $^{58,64}$Ni+$^{208}$Pb
reactions examined in the present paper.

\subsection{For the $^{64}$Ni+$^{208}$Pb system}

\begin{figure} [tb]
\includegraphics[width=8cm]{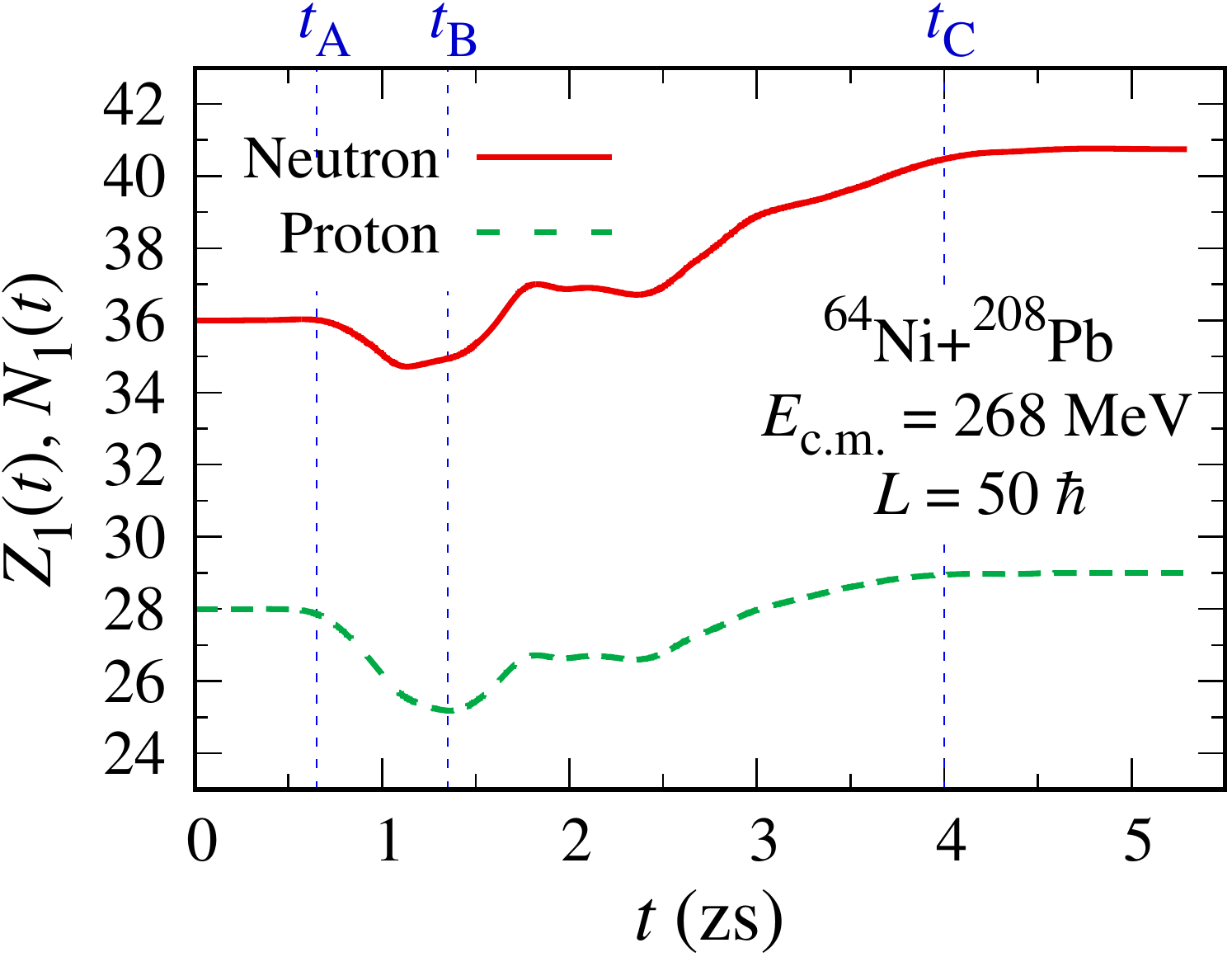}\vspace{-1mm}
\caption{
The smoothed mean values of the neutron and proton numbers of the projectile-like subsystem
in the $^{64}$Ni+$^{208}$Pb reaction at $E_{\rm c.m.}$\,$=$\,268~MeV with the initial orbital
angular momentum $L$\,$=$\,50$\hbar$ as a function of time. Red solid line shows that of neutrons,
$N_1(t)$, while the green dashed line shows that of protons, $Z_1(t)$.
}\vspace{-3mm}
\label{App:fig1}
\end{figure}

From a TDHF calculation, one can compute time evolution of the mean neutron and proton numbers
in the projectile-like subsystem, $N_1(t)$ and $Z_1(t)$. Because of the complex fluctuations in the window
position as well as shell structure, those quantities show rapid fluctuations as a function of time.
In order to eliminate the rapid fluctuations in time, we carry out a smoothing by taking an average
over a short time interval, $T_{\rm ave}\simeq$\,0.67~zs. In Fig.~\ref{App:fig1}, we show the
smoothed mean neutron and proton numbers in the projectile-like subsystem, $N_1(t)$ (red solid line) and
$Z_1(t)$ (green dashed line), in the $^{64}$Ni+$^{208}$Pb reaction at $E_{\rm c.m.}$\,$=$\,268~MeV
with the initial orbital angular momentum $L$\,$=$\,50$\hbar$ as a function of time. One can see that
there is a fast charge equilibration process (from $t_{\rm A}$ to $t_{\rm B}$), followed by a slow
mass equilibration process (from $t_{\rm B}$ to $t_{\rm C}$), during the collision process.

\begin{figure} [tb]
\includegraphics[width=8.6cm]{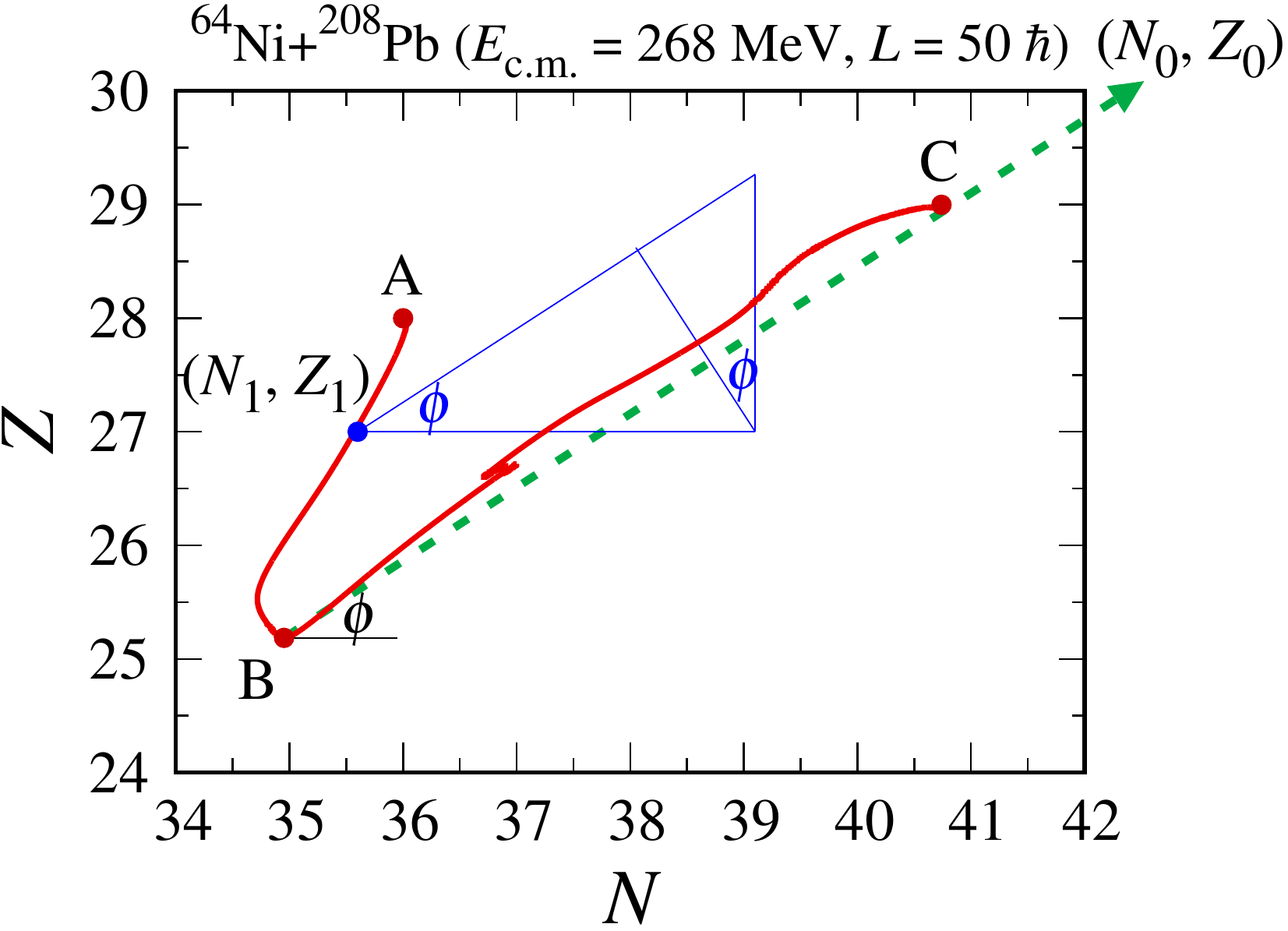}\vspace{-1mm}
\caption{
The smoothed mean drift path in the $N$-$Z$ plane for the projectile-like subsystem
in the $^{64}$Ni+$^{208}$Pb reaction at $E_{\rm c.m.}$\,$=$\,268~MeV with the initial
orbital angular momentum $L$\,$=$\,50$\hbar$. Green dashed arrow indicates the isoscalar
path which continues along the mass symmetry point, $(N_0,Z_0)=(81,55)$. The angle
$\phi$\,$=$\,32.9$^\circ$ is the angle between the isoscalar path and the neutron axis.
The points A, B, and C correspond to the time $t_{\rm A}$, $t_{\rm B}$, and
$t_{\rm C}$ in Fig.~\ref{App:fig1}, respectively. Blue triangles help to understand
the distances, Eqs.~(\ref{Eq:R_S}) and (\ref{Eq:R_E}).
}\vspace{-3mm}
\label{App:fig2}
\end{figure}

In Fig.~\ref{App:fig2}, the same quantities as shown in Fig.~\ref{App:fig1} are shown,
but now they are plotted in the $N$-$Z$ plane, which we call the drift path. In the figure,
green dashed line indicates a collection of nuclei with nearly same charge asymmetry values,
$\delta$\,$=$\,$(N-Z)/(N+Z)$\,$=$\,0.16--0.19. We call this line the isoscalar path which extends
along the beta-stability valley until the mass equilibrium at $N_0$\,$=$\,(36\,+\,126)/2\,$=$\,81
and $Z_0$\,$=$\,(28\,+\,82)/2\,$=$\,55. Starting from the point A, the ($N_1,Z_1$) follows
the red solid line until it reaches the charge equilibrium at $\delta$\,$\simeq$\,0.16 at
the point B and drifts toward the mass symmetry nearly along the isoscalar path. Note
that the binary system separates before reaching the mass equilibrium of the system.

The potential energy surface of the dinuclear system with respect to ($N_1,Z_1$) provided
by the microscopic Skyrme energy density functional would have a rather complex structure.
Here, we approximate the potential by a two-parabolic form: One parabola is along
the isoscalar direction and the other is perpendicular to it, which we call the isovector path.
Namely, the driving potential is expressed as
\begin{equation}
U(N_1,Z_1) = \frac{1}{2}aR_S^2 + \frac{1}{2}bR_E^2,
\label{Eq:PES}
\end{equation}
where
\begin{eqnarray}
R_S &=& \bigl[ N_0 - N_1 \bigr]\sin\phi - \bigl[Z_0-Z_1\bigr]\cos\phi,\;\;\label{Eq:R_S}\\
R_E &=& \bigl[ N_0 - N_1 \bigr]\cos\phi + \bigl[Z_0-Z_1\bigr]\sin\phi.\;\;\label{Eq:R_E}
\end{eqnarray}
$R_S(t)$ represents the distance of ($N_1,Z_1$) on the drift path from the isoscalar
path, while $R_E(t)$ represents the distance of ($N_1,Z_1$) from the equilibrium point
($N_0,Z_0$) along the isoscalar path. The angle, $\phi$\,$=$\,32.9$^\circ$, denotes
the angle between the isoscalar path and the neutron axis. The Einstein relation relates
the drift coefficients to the derivatives of the potential energy surface according to
\begin{eqnarray}
\nu_n &=& -\frac{D_{NN}}{T^*}\frac{\partial}{\partial N_1}U(N_1,Z_1),\\
\nu_p &=& -\frac{D_{ZZ}}{T^*}\frac{\partial}{\partial Z_1}U(N_1,Z_1).
\end{eqnarray}
Using the two-parabolic form of the potential (\ref{Eq:PES}), one finds analytical
expressions of the drift coefficients:
\begin{eqnarray}
\nu_n(t) &=& +D_{NN}(t)\bigl[ \alpha R_S(t)\sin\phi + \beta R_E(t)\cos\phi \bigr],\label{Eq:nu_n}\\
\nu_p(t) &=& -D_{ZZ}(t)\bigl[ \alpha R_S(t)\cos\phi - \beta R_E(t)\sin\phi \bigr].\label{Eq:nu_p}
\end{eqnarray}
Note that the effective temperature $T^*$ has been absorbed to the curvature parameters,
$\alpha=a/T^*$ and $\beta=b/T^*$. Having those analytical forms at hand, one can
readily calculate the derivatives of the drift coefficients, i.e.,
\begin{eqnarray}
\frac{\partial\nu_n(t)}{\partial N_1} &=& -D_{NN}(t) \bigl[ \alpha\sin^2\phi + \beta\cos^2\phi \bigr],\label{Eq:dnu_NN}\\
\frac{\partial\nu_p(t)}{\partial Z_1} &=& -D_{ZZ}(t) \bigl[ \alpha\cos^2\phi + \beta\sin^2\phi \bigr],\label{Eq:dnu_ZZ}\\
\frac{\partial\nu_n(t)}{\partial Z_1} &=& D_{NN}(t)(\alpha-\beta)\sin\phi\cos\phi,\label{Eq:dnu_NZ}\\
\frac{\partial\nu_p(t)}{\partial N_1} &=& D_{ZZ}(t)(\alpha-\beta)\sin\phi\cos\phi.\label{Eq:dnu_ZN}
\end{eqnarray}

The remaining task is to determine the reduced curvature parameters, $\alpha$ and $\beta$.
Actually, the latter quantities can be determined with the mean drift coefficients, $\nu_n(t)=
dN_1(t)/dt$ and $\nu_p(t)=dZ_1(t)/dt$, obtained from TDHF. From Eqs.~(\ref{Eq:nu_n})
and (\ref{Eq:nu_p}), we have the following equalities:
\begin{eqnarray}
\alpha\,R_S(t) &=& \frac{\nu_n(t)\sin\phi}{D_{NN}(t)} - \frac{\nu_p(t)\cos\phi}{D_{ZZ}(t)},\\
\beta\,R_E(t) &=& \frac{\nu_n(t)\cos\phi}{D_{NN}(t)} + \frac{\nu_p(t)\sin\phi}{D_{ZZ}(t)}.
\end{eqnarray}
To derive macroscopic drift coefficients for the nucleon diffusion mechanism, we need to
eliminate microscopic effects on the potential energy, e.g., complex dynamic shell effects.
Namely, due to the complex structure of the microscopic potential energy surface in TDHF,
the reduced curvature parameters of the simple parabolic approximation may vary in time.
We thus take the following time average to determine the average reduced curvature parameters:
\begin{eqnarray}
\bar{\alpha} &=&
\int_{t_A}^{t_B} \biggl( \frac{\nu_n(t)\sin\phi}{D_{NN}(t)} - \frac{\nu_p(t)\cos\phi}{D_{ZZ}(t)} \biggr)dt
\,\bigg/\int_{t_A}^{t_B}\hspace{-2mm} R_S(t)\,dt,\label{Eq:alpha_ave}\nonumber\\\\
\bar{\beta} &=& 
\int_{t_B}^{t_C} \biggl( \frac{\nu_n(t)\cos\phi}{D_{NN}(t)} + \frac{\nu_p(t)\sin\phi}{D_{ZZ}(t)} \biggr)dt
\,\bigg/\int_{t_B}^{t_C}\hspace{-2mm} R_E(t)\,dt.\label{Eq:beta_ave}\nonumber\\
\end{eqnarray}
Note that all quantities in the right-hand side of Eqs.~(\ref{Eq:alpha_ave}) and (\ref{Eq:beta_ave})
can be computed within the TDHF approach. Here, $t_{\rm A}$\,$=$\,0.65~zs, $t_{\rm B}$\,$=$\,1.35~zs,
and $t_{\rm C}$\,$=$\,4.00~zs indicate the times at the points A, B, and C, respectively, in Figs.~\ref{App:fig1}
and \ref{App:fig2}. For the $^{64}$Ni+$^{208}$Pb system, we obtained the
average reduced curvature parameters of $\bar{\alpha}$\,$=$\,0.289 and $\bar{\beta}$\,$=$\,0.003
for the isovector and isoscalar directions, respectively. With those $\bar{\alpha}$ and $\bar{\beta}$,
the derivatives of the drift coefficients, Eqs.~(\ref{Eq:dnu_NN})--(\ref{Eq:dnu_ZN}), were evaluated.

\subsection{For the $^{58}$Ni+$^{208}$Pb system}

\begin{figure} [t]
\includegraphics[width=8cm]{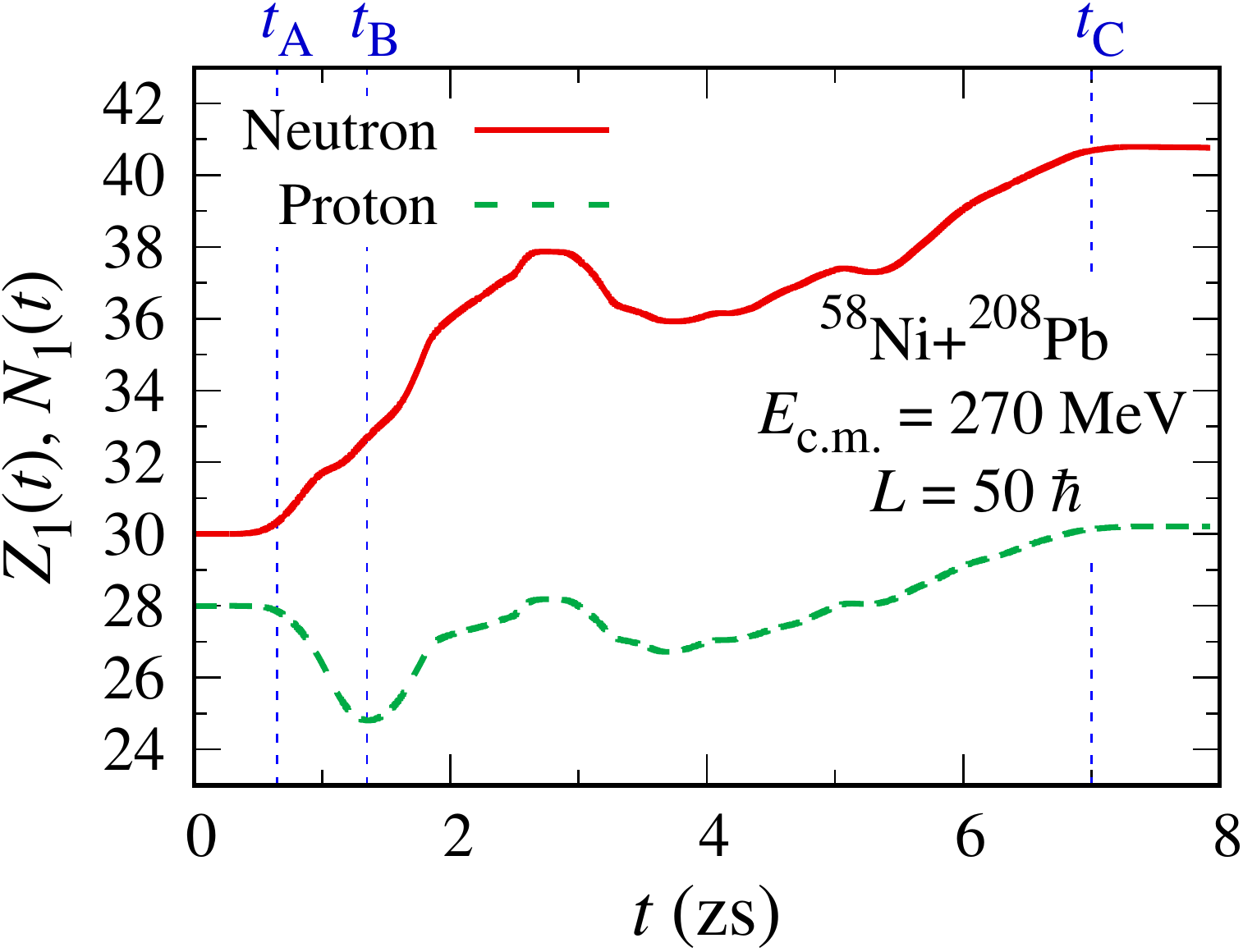}\vspace{-1mm}
\caption{
Same as Fig.~\ref{App:fig1}, but for the $^{58}$Ni+$^{208}$Pb reaction at
$E_{\rm c.m.}$\,$=$\,270~MeV with the initial orbital angular momentum $L$\,$=$\,50$\hbar$.
}\vspace{-3mm}
\label{App:fig3}
\end{figure}

\begin{figure} [t]
\includegraphics[width=8.6cm]{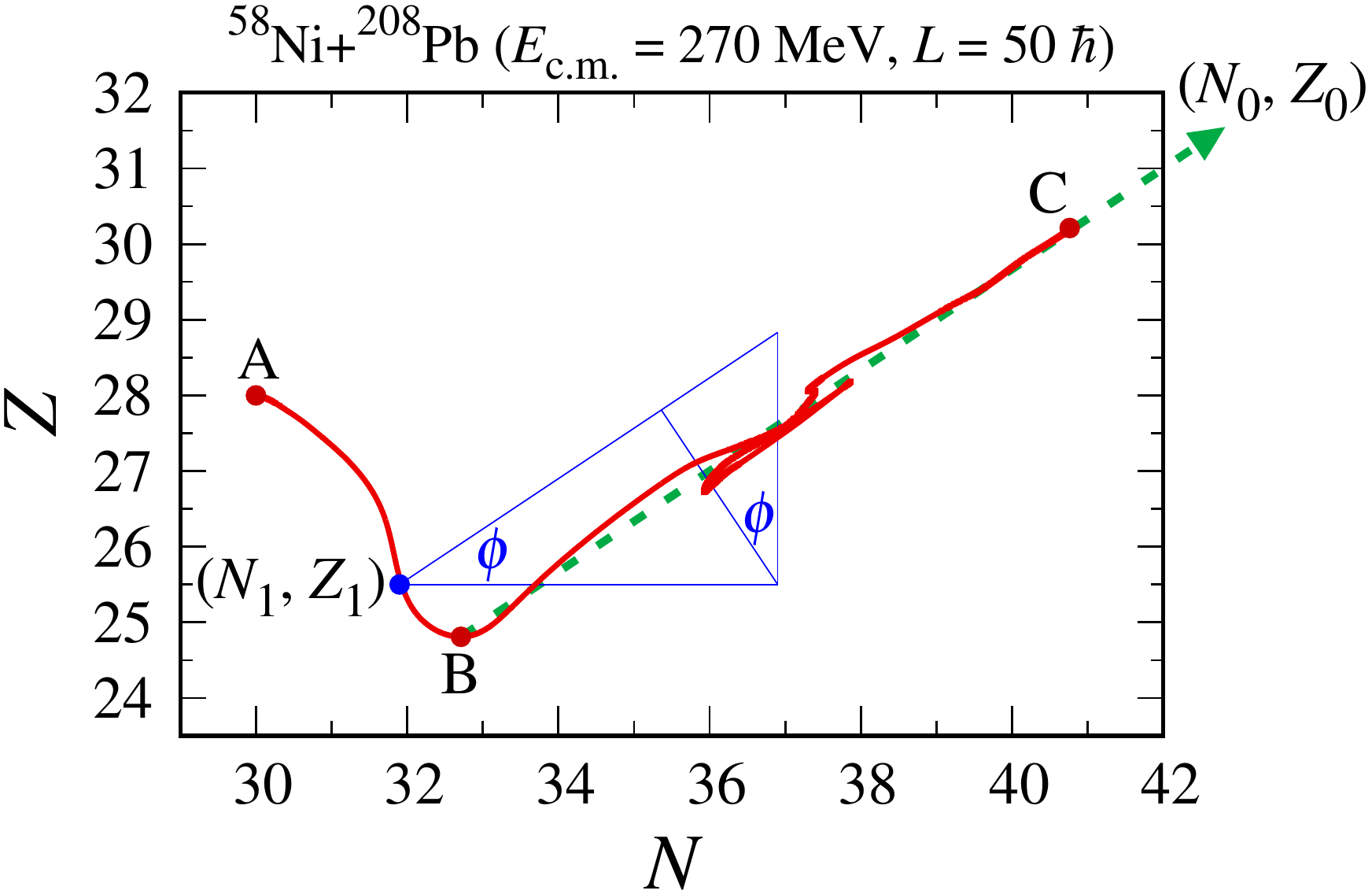}\vspace{-1mm}
\caption{
Same as Fig.~\ref{App:fig2}, but for the $^{58}$Ni+$^{208}$Pb reaction at
$E_{\rm c.m.}$\,$=$\,270~MeV with the initial orbital angular momentum $L$\,$=$\,50$\hbar$.
The mass symmetry point is $(N_0,Z_0)=(78,55)$ and the angle between the isoscalar path
and the neutron axis is $\phi$\,$=$\,33.3$^\circ$.
}\vspace{-3mm}
\label{App:fig4}
\end{figure}

Figures~\ref{App:fig3} and \ref{App:fig4} show the time evolution of the smoothed mean
neutron and proton numbers in the projectile-like subsystem in the $^{58}$Ni+$^{208}$Pb
reaction at $E_{\rm c.m.}$\,$=$\,270~MeV with the initial orbital angular momentum $L$\,$=$\,50$\hbar$.

Basically, we repeat the same procedure also for the $^{58}$Ni+$^{208}$Pb system.
The times $t_{\rm A}$\,$=$\,0.65~zs, $t_{\rm B}$\,$=$\,1.35~zs, and $t_{\rm C}$\,$=$\,7.00~zs
of the averaging intervals indicated in Fig.~\ref{App:fig3} correspond to the points A, B, and C
in Fig.~\ref{App:fig4}, respectively. In the isoscalar drift path of this system between the points
B and C in Fig.~\ref{App:fig4}, the nuclei have nearly the same charge asymmetry values
of $\delta$\,$=$\,0.14--0.17. The angle between the isoscalar path and the neutron axis is
$\phi$\,$=$\,33.3$^\circ$, which is similar to the one for the $^{64}$Ni+$^{208}$Pb system.
The equilibrium values of neutron and proton numbers are $N_0$\,$=$\,(30\,+\,126)/2\,$=$\,78
and $Z_0$\,$=$\,(28\,+\,82)/2\,$=$\,55. The average curvature parameters evaluated by
Eqs.~(\ref{Eq:alpha_ave}) and (\ref{Eq:beta_ave}) have nearly the same magnitude as
those for the $^{64}$Ni+$^{208}$Pb system, which are given by $\bar{\alpha}$\,$=$\,0.247
and $\bar{\beta}$\,$=$\,0.002 for the isovector and isoscalar directions, respectively.

We note that the isoscalar line (green dashed arrow) in Figs.~\ref{App:fig2} and
\ref{App:fig4} should be extended further until the mass equilibrium point, $(N_0,Z_0)$.

\section{Tables of the TDHF and SMF results}\label{Appendix:Tables}

Here, we provide the numerical results of the TDHF and SMF calculations for the $^{64}$Ni+$^{208}$Pb reaction
at $E_{\rm c.m.}$\,$=$\,268~MeV in Table~\ref{Table:64Ni+208Pb} and for the $^{58}$Ni+$^{208}$Pb
reaction at $E_{\rm c.m.}$\,$=$\,270~MeV in Table~\ref{Table:58Ni+208Pb} with various initial conditions.

\begin{table*}[t]
\caption{
Results of the TDHF and SMF calculations for the $^{64}$Ni+$^{208}$Pb reaction at $E_{\rm c.m.}$\,$=$\,268~MeV.
From left to right columns, the table lists: the initial orbital angular momentum $L_i$ in $\hbar$, the corresponding
impact parameter $b$ in fm, the final values of the average proton and neutron numbers in a projectile-like fragment
(PLF) ($Z_1^f$ and $N_1^f$) and a target-like fragment (TLF) ($Z_2^f$ and $N_2^f$), the final orbital angular momentum
$L_f$ in $\hbar$, the total kinetic energy loss (TKEL) in MeV, the contact time $t_{\rm contact}$ in zeptoseconds
(1\,zs\,$=$\,10$^{-21}$\,s), the dispersions by SMF ($\sigma_{NN}$, $\sigma_{ZZ}$, $\sigma_{NZ}$, and
$\sigma_{AA}$) and the mass dispersion by TDHF ($\sigma_{AA}$), and the scattering angles in the center-of-mass
frame ($\theta_{\rm c.m.}$) and in the laboratory frame for a PLF ($\vartheta_1^{\rm lab}$) and a TLF
($\vartheta_2^{\rm lab}$) in degrees.
}
\label{Table:64Ni+208Pb}
\begin{center}
\begin{tabular*}{\textwidth}{@{\extracolsep{\fill}}ccccccccccccccccc}
\hline\hline
$L_i$ ($\hbar$) & $b$ (fm) & $Z_1^f$ & $N_1^f$ & $Z_2^f$ & $N_2^f$ & $L_f$ ($\hbar$) & TKEL (MeV) & $t_{\rm contact}$ (zs) & $\sigma_{NN}$ & $\sigma_{ZZ}$ & $\sigma_{NZ}$ & $\sigma_{AA}$ & $\sigma_{AA}^{\rm TDHF}$ & $\theta_{\rm c.m.}$ & $\vartheta_1^{\rm lab}$ & $\vartheta_2^{\rm lab}$ (deg) \\
\hline
0	&0.00	&27.75	&37.99	&82.23	&123.81	&0.0 	&64.1	&4.66	&11.98	&7.85	&9.50	&19.64	&1.94	&180.0	&180.0	&0.0\\
5	&0.20	&27.73	&37.98	&82.24	&123.83	&4.3 	&64.3	&4.62	&11.92	&7.81	&9.45	&19.53	&1.94	&169.9	&164.3	&4.7\\
10	&0.40	&27.69	&37.95	&82.29	&123.85	&8.4 	&65.2	&4.56	&11.85	&7.77	&9.40	&19.43	&1.94	&160.2	&149.8	&9.3\\
15	&0.60	&27.60	&37.88	&82.37	&123.91	&12.8	&66.9	&4.45	&11.80	&7.74	&9.36	&19.34	&1.94	&150.8	&136.4	&13.7\\
20	&0.80	&27.49	&37.79	&82.48	&123.99	&17.4	&68.7	&4.36	&11.74	&7.70	&9.30	&19.24	&1.94	&141.8	&124.4	&17.8\\
25	&0.99	&27.48	&37.87	&82.49	&123.92	&21.0	&70.7	&4.37	&11.30	&7.41	&8.94	&18.50	&1.94	&133.2	&113.8	&21.7\\
30	&1.19	&27.72	&38.25	&82.26	&123.55	&23.5	&71.5	&4.35	&11.24	&7.37	&8.89	&18.41	&1.96	&125.4	&104.7	&25.3\\
35	&1.39	&28.14	&38.83	&81.83	&122.99	&25.1	&70.0	&4.30	&11.16	&7.32	&8.83	&18.27	&1.97	&118.5	&97.1	&28.7\\
40	&1.59	&28.40	&39.34	&81.57	&122.50	&27.1	&67.9	&4.10	&11.01	&7.23	&8.71	&18.03	&1.97	&112.4	&90.7	&31.7\\
45	&1.79	&28.62	&39.86	&81.36	&121.99	&33.3	&65.0	&3.82	&10.76	&7.06	&8.50	&17.61	&1.98	&106.7	&85.1	&34.6\\
50	&1.99	&28.99	&40.69	&80.99	&121.18	&37.3	&62.3	&3.57	&10.41	&6.83	&8.20	&17.02	&1.91	&104.4	&82.8	&36.1\\
55	&2.19	&29.31	&41.14	&80.67	&120.73	&41.6	&61.8	&3.23	&10.00	&6.56	&7.86	&16.33	&1.91	&102.8	&81.1	&37.1\\
60	&2.39	&29.33	&41.09	&80.66	&120.77	&46.0	&65.1	&3.02	&9.67	&6.36	&7.60	&15.79	&1.92	&99.8	&78.1	&38.3\\
65	&2.59	&29.27	&40.94	&80.72	&120.93	&49.6	&66.3	&2.82	&9.34	&6.13	&7.31	&15.23	&1.94	&96.7	&75.3	&39.6\\
70	&2.79	&29.19	&40.86	&80.80	&121.03	&53.8	&64.1	&2.54	&8.95	&5.88	&6.99	&14.58	&1.91	&94.2	&73.2	&40.9\\
75	&2.98	&29.12	&40.99	&80.87	&120.91	&59.2	&63.1	&2.34	&8.56	&5.63	&6.66	&13.92	&1.88	&91.8	&71.1	&42.1\\
80	&3.18	&29.26	&41.15	&80.73	&120.76	&63.9	&62.6	&2.15	&8.16	&5.37	&6.33	&13.25	&1.86	&90.5	&69.9	&42.8\\
85	&3.38	&29.29	&40.73	&80.70	&121.18	&67.5	&62.3	&1.96	&7.74	&5.10	&5.98	&12.55	&1.85	&90.0	&69.6	&42.9\\
90	&3.58	&28.96	&39.90	&81.04	&122.03	&70.7	&62.3	&1.79	&7.31	&4.83	&5.61	&11.82	&1.85	&89.8	&69.6	&42.7\\
95	&3.78	&28.61	&39.05	&81.38	&122.86	&76.9	&65.7	&1.69	&6.85	&4.53	&5.22	&11.04	&1.85	&88.6	&68.6	&42.7\\
100	&3.98	&27.89	&37.84	&82.10	&124.07	&80.8	&70.0	&1.48	&6.28	&4.18	&4.74	&10.09	&1.80	&88.5	&68.7	&41.8\\
110	&4.38	&26.92	&36.81	&83.06	&125.10	&86.9	&66.9	&1.02	&5.16	&3.45	&3.74	&8.16 	&1.68	&91.4	&71.7	&40.2\\
120	&4.77	&26.77	&36.69	&83.22	&125.23	&100.1	&54.0	&0.66	&4.12	&2.79	&2.78	&6.34 	&1.50	&92.8	&73.6	&40.4\\
130	&5.17	&27.30	&36.45	&82.70	&125.52	&118.5	&29.9	&0.02	&2.85	&1.79	&1.38	&3.89 	&1.16	&93.5	&75.1	&41.6\\
140	&5.57	&27.86	&36.20	&82.14	&125.80	&137.0	&6.1 	&0.00	&1.72	&0.92	&0.41	&2.03 	&0.74	&93.2	&75.7	&43.1\\
150	&5.97	&27.96	&36.10	&82.04	&125.90	&148.8	&2.3 	&0.00	&1.20	&0.64	&0.20	&1.39 	&0.52	&90.5	&73.3	&44.6\\
160	&6.37	&27.98	&36.06	&82.02	&125.94	&159.4	&1.3 	&0.00	&0.92	&0.50	&0.12	&1.06 	&0.40	&87.4	&70.5	&46.3\\
170	&6.76	&27.99	&36.03	&82.01	&125.96	&169.7	&0.9 	&0.00	&0.73	&0.40	&0.08	&0.84 	&0.32	&84.2	&67.6	&47.9\\
180	&7.16	&28.00	&36.02	&82.00	&125.98	&179.9	&0.7 	&0.00	&0.60	&0.33	&0.05	&0.69 	&0.26	&81.1	&64.9	&49.4\\
190	&7.56	&28.00	&36.01	&82.00	&125.98	&189.9	&0.5 	&0.00	&0.50	&0.27	&0.03	&0.57 	&0.21	&78.2	&62.3	&50.9\\
200	&7.96	&28.00	&36.01	&82.00	&125.99	&200.0	&0.4 	&0.00	&0.42	&0.22	&0.02	&0.48 	&0.17	&75.4	&59.9	&52.3\\
210	&8.36	&28.00	&36.01	&82.00	&125.99	&210.0	&0.4 	&0.00	&0.36	&0.18	&0.02	&0.40 	&0.13	&72.7	&57.7	&53.6\\
220	&8.75	&28.00	&36.00	&82.00	&126.00	&220.0	&0.3 	&0.00	&0.31	&0.15	&0.01	&0.34 	&0.11	&70.2	&55.5	&54.9\\
230	&9.15	&28.00	&36.00	&82.00	&126.00	&230.0	&0.3 	&0.00	&0.26	&0.12	&0.01	&0.29 	&0.09	&67.9	&53.6	&56.0\\
240	&9.55	&28.00	&36.00	&82.00	&126.00	&240.0	&0.3 	&0.00	&0.23	&0.10	&0.01	&0.25 	&0.07	&65.7	&51.7	&57.1\\
\hline\hline
\end{tabular*}\vspace{-2mm}
\end{center}
\end{table*}

\begin{table*}[t]
\caption{
Same as Table~\ref{Table:64Ni+208Pb}, but for the $^{58}$Ni+$^{208}$Pb reaction at $E_{\rm c.m.}$\,$=$\,270~MeV.
}
\label{Table:58Ni+208Pb}
\begin{center}
\begin{tabular*}{\textwidth}{@{\extracolsep{\fill}}ccccccccccccccccc}
\hline\hline
$L_i$ ($\hbar$) & $b$ (fm) & $Z_1^f$ & $N_1^f$ & $Z_2^f$ & $N_2^f$ & $L_f$ ($\hbar$) & TKEL (MeV) & $t_{\rm contact}$ (zs) & $\sigma_{NN}$ & $\sigma_{ZZ}$ & $\sigma_{NZ}$ & $\sigma_{AA}$ & $\sigma_{AA}^{\rm TDHF}$ & $\theta_{\rm c.m.}$ & $\vartheta_1^{\rm lab}$ & $\vartheta_2^{\rm lab}$ (deg) \\
\hline
50	&2.06	&30.20	&40.70	&79.77	&115.12	&36.7	&67.8	&6.21	&13.42	&9.04	&10.81	&22.26	&2.10	&57.8	&43.2	&60.5\\
55	&2.26	&29.78	&39.57	&80.20	&116.32	&40.6	&61.4	&4.38	&11.63	&7.84	&9.31	&19.24	&2.03	&77.0	&59.1	&51.1\\
60	&2.47	&29.37	&39.07	&80.61	&116.84	&43.4	&62.0	&3.11	&9.75	&6.58	&7.73	&16.06	&2.00	&92.1	&72.4	&43.3\\
65	&2.68	&29.28	&39.18	&80.70	&116.74	&50.1	&64.1	&2.71	&8.98	&6.07	&7.08	&14.76	&1.98	&91.8	&72.0	&43.4\\
70	&2.88	&28.73	&37.99	&81.25	&117.92	&53.6	&71.7	&2.36	&8.35	&5.65	&6.54	&13.68	&2.07	&90.8	&71.1	&42.8\\
75	&3.09	&28.44	&36.85	&81.54	&119.08	&60.3	&70.9	&2.01	&7.52	&5.10	&5.83	&12.27	&2.04	&92.6	&73.0	&41.6\\
80	&3.29	&27.32	&35.17	&82.66	&120.74	&58.8	&75.3	&1.57	&6.52	&4.44	&4.96	&10.56	&1.97	&97.6	&77.9	&38.3\\
85	&3.50	&26.42	&34.04	&83.56	&121.87	&63.6	&73.9	&1.19	&5.57	&3.82	&4.11	&8.92 	&1.82	&99.9	&80.5	&36.9\\
90	&3.71	&25.78	&33.17	&84.20	&122.74	&73.1	&70.7	&0.70	&4.85	&3.33	&3.43	&7.63 	&1.76	&99.9	&81.0	&36.7\\
95	&3.91	&26.00	&33.16	&83.99	&122.77	&80.2	&59.1	&0.49	&4.21	&2.90	&2.80	&6.47 	&1.66	&101.1	&82.6	&36.9\\
100	&4.12	&26.21	&32.73	&83.79	&123.22	&88.9	&44.1	&0.29	&3.54	&2.42	&2.10	&5.21 	&1.46	&101.9	&84.0	&37.2\\
110	&4.53	&27.48	&31.80	&82.52	&124.19	&106.0	&12.6	&0.00	&2.24	&1.32	&0.74	&2.81 	&1.10	&102.3	&85.6	&38.6\\
120	&4.94	&27.84	&30.93	&82.16	&125.07	&117.8	& 4.1	&0.00	&1.58	&0.85	&0.33	&1.86 	&0.88	&99.9	&83.6	&40.1\\
130	&5.35	&27.93	&30.56	&82.07	&125.44	&128.3	& 2.4	&0.00	&1.23	&0.66	&0.20	&1.43 	&0.72	&96.2	&80.1	&41.9\\
140	&5.76	&27.96	&30.35	&82.04	&125.65	&138.6	& 1.8	&0.00	&0.99	&0.54	&0.13	&1.14 	&0.58	&92.4	&76.6	&43.8\\
150	&6.18	&27.98	&30.23	&82.02	&125.77	&148.8	& 1.4	&0.00	&0.80	&0.45	&0.09	&0.93 	&0.48	&88.7	&73.2	&45.6\\
160	&6.59	&27.99	&30.15	&82.01	&125.85	&159.0	& 1.2	&0.00	&0.65	&0.38	&0.06	&0.76 	&0.39	&85.2	&69.9	&47.4\\
170	&7.00	&27.99	&30.09	&82.01	&125.90	&169.1	& 1.0	&0.00	&0.53	&0.32	&0.04	&0.63 	&0.31	&81.8	&66.9	&49.1\\
180	&7.41	&28.00	&30.06	&82.00	&125.94	&179.2	& 0.9	&0.00	&0.44	&0.27	&0.03	&0.52 	&0.25	&78.7	&64.1	&50.6\\
190	&7.82	&28.00	&30.04	&82.00	&125.96	&189.3	& 0.8	&0.00	&0.36	&0.23	&0.02	&0.43 	&0.20	&75.7	&61.5	&52.1\\
200	&8.23	&28.00	&30.02	&82.00	&125.97	&199.3	& 0.7	&0.00	&0.30	&0.19	&0.01	&0.36 	&0.16	&72.9	&59.1	&53.5\\
210	&8.65	&28.00	&30.02	&82.00	&125.98	&209.4	& 0.7	&0.00	&0.25	&0.16	&0.01	&0.30 	&0.13	&70.3	&56.8	&54.8\\
220	&9.06	&28.00	&30.01	&82.00	&125.99	&219.4	& 0.6	&0.00	&0.21	&0.14	&0.01	&0.25 	&0.10	&67.8	&54.7	&56.0\\
230	&9.47	&28.00	&30.01	&82.00	&125.99	&229.5	& 0.6	&0.00	&0.17	&0.12	&0.00	&0.21 	&0.08	&65.5	&52.7	&57.2\\
240	&9.88	&28.00	&30.00	&82.00	&126.00	&239.5	& 0.5	&0.00	&0.15	&0.10	&0.00	&0.18 	&0.07	&63.3	&50.8	&58.3\\
\hline\hline
\end{tabular*}\vspace{-2mm}
\end{center}
\end{table*}

\end{document}